\documentstyle[aps,pre,multicol]{revtex}
\begin{document}
\def\be{\begin{equation}}
\def\ee{\end{equation}}
\def\bc{\begin{center}}
\def\ec{\end{center}}
\def\bea{\begin{eqnarray}}
\def\eea{\end{eqnarray}}

\title{Condensation vs. phase-ordering in the dynamics of first
order transitions}
\author{C. Castellano$^{1}$, F. Corberi$^{1,2}$ and M. Zannetti$^{1}$}
\address {$1$ Istituto Nazionale di Fisica della Materia, Unit\`{a}
di Salerno and Dipartimento di Fisica, Universit\`{a} di Salerno,
84081 Baronissi (Salerno), Italy}
\address {$2$ Dipartimento di Scienze Fisiche, Universit\`{a}
di Napoli, Mostra d'Oltremare, Padiglione 19, 80125 Napoli, Italy}
\maketitle
\begin{abstract}
The origin of the non commutativity of the limits $t \rightarrow \infty$
and $N \rightarrow \infty$ in the dynamics of first order transitions is
investigated. In the large-$N$ model, i.e. $N \rightarrow \infty$ taken first,
the low temperature phase is characterized by condensation of the
large wave length fluctuations rather than by genuine phase-ordering
as when $t \rightarrow \infty$ is taken first. A detailed study of
the scaling properties of the structure factor in the large-$N$ 
model is carried out for quenches above, at and below $T_c$.
Preasymptotic scaling is found and crossover phenomena are related
to the existence of components in the order parameter with 
different scaling properties. Implications for phase-ordering
in realistic systems are discussed.
\end{abstract}

\noindent e-mail: castellano@na.infn.it, corberi@na.infn.it,
zannetti@na.infn.it

%\begin{multicols}{2}
\section{Introduction}
The large time behavior of a system quenched at or
below the critical point is characterized by scale invariance~\cite{Bray94}.
For the equal time structure factor one has
\be
C({\vec k},t) = L^{\alpha}(t) F(k L(t))
\label{SS}
\ee
where
\be
L(t) \sim t^{1 \over z}
\label{Growthlaw}
\ee
is the characteristic length growing with time according to a power law
and $F(x)$ is the scaling function.
The physics behind~(\ref{SS}) is quite simple and is
basically due to the degeneracy of the low temperature state.
After the quench an exponentially fast process
takes place leading to local equilibrium.
If multiple choice is available, correlated regions of the possible
low temperature phases are formed.
From that point onward equilibration proceeds through
the coarsening of these correlated regions, whose characteristic size
$L(t)$ grows according to~(\ref{Growthlaw}).
The difference between quenches to $T_c$ and below $T_c$ is that 
in the first case the correlated regions are fractal (Appendix I), while in
the second one are compact.
Apart from this, in both cases the equilibration process becomes slow
(if the system is infinite, equilibrium is never reached) and
after domains of the ordered phases have formed, 
scaling behavior occurs since the residual time dependence is confined
in the typical size $L(t)$ of the correlated regions.

The whole time evolution can be divided into a preasymptotic
and  an asymptotic regime, with a smooth transition between the two.
The asymptotic regime displays universality and it is controlled by a fixed
point structure.
The universality classes are determined by features like the presence
or absence of a conservation law, the number $N$ of components of the order
parameter, the dimensionality $d$ of space and the final 
temperature $T_F$ of the quench.
More precisely, on the temperature axis there is an unstable fixed point
at the critical temperature $T_c$ and an attractive fixed point at $T_F=0$.
For the exponent $\alpha$ one has
\be
\alpha = \left \{
\begin{array}{cc}
2-\eta, &\mbox{for } T_F = T_c \\
d, &\mbox{for } T_F < T_c
\end{array}
\right.
\label{eqalpha}
\ee
where $\eta$ is the usual exponent of the static critical phenomena.
The exponent $z$ coincides
with the exponent of the dynamical critical phenomena for $T_F=T_c$.
Instead, for any final temperature below $T_c$, $z=2$ for non conserved
order parameter (NCOP), while for
conserved order parameter (COP) $z=3$ when $N=1$ and $z=4$ when $N>1$.
The scaling function $F(x)$ also displays universal features and it is
sensitive to the space dimensionality through the presence ($N<d$)
or absence ($N>d$) of localized topological defects.
By contrast, in the preasymptotic regime the evolution of the system
is not universal, as it depends on the initial conditions of the
quench and on the actual value of the final temperature.

A complete theory of the process then should derive the scaling
behavior from the basic equation of motion for the order parameter
and should be able to describe how the relatively simple universal
asymptotic regime emerges out of the complexity of the preasymptotic
regime.
Ideally, one would like to have a manageable reference theory
which accounts, at least qualitatively, for the basic features of the process
and a systematic procedure for the computation of 
the corrections~\cite{Mazenko88}.
A scheme of this type is available for quenches to $T_c$,
where, despite the difficulty due to the lack of time translational
invariance, the field theoretical machinery developed for critical
phenomena is to a large extent applicable~\cite{Janssen89}.
Instead, for quenches below $T_c$ the present status of theoretical
understanding is far from this standard.
What we have in this case is the linear theory~\cite{Cahn59} for the very early
stage of the process, which applies only when initial conditions are so
small that it is actually justified to employ a linear approximation, and
{\it ad hoc} late stage theories~\cite{Ohta82}.
Although these late stage theories have had much success in the computation
of the scaling functions, yet are based on uncontrolled approximations.
Furthermore, the late stage theories do not connect to the early stage theory,
if this is available at all. So, there is no theoretical understanding of
the complex phenomenology arising at the breakdown of the early stage
theory and leading to the onset of scaling~\cite{Langer92}.
Proposals for the systematic improvement of the late stage theory have been
put forward~\cite{Bray93} but, as of now, a first principles 
theory of phase-ordering processes is out of reach.

In this theoretical landscape a special position is occupied by the
$1/N$-expansion.
As applied to critical phenomena this technique provides a very clear
instance of what is to be understood for a systematic theory: there is
a lowest order analytically tractable approximation
(the large-$N$ model) which captures the basic physics and there is
an expansion parameter ($1/N$) which allows for the systematic computation
of the corrections.
The scheme applies succesfully also to quenches 
to the critical point~\cite{Janssen89} and,
at first sight, it would seem to be applicable as well to the phase-ordering
processes.
Indeed, in the large-$N$ model one can solve exactly~\cite{Coniglio89} 
for the structure factor
and one finds that the standard scaling form (\ref{SS}) is obeyed
for large time with NCOP~\cite{Mazenko85}.
In particular one finds $z=2$, and $\alpha$ is given by (3) with $\eta=0$.
The scaling functions can also be found explicitely~\cite{Coniglio94}.
It is to be stressed that in the solution of the model with $N=\infty$
there are no {\it ad hoc} hypotheses and the above outlined picture of the
asymptotic behavior with scaling, universality
and temperature fixed points is derived from the solution
of the equation of motion.

However, when the model is solved with COP~\cite{Coniglio89,Coniglio94}, 
although the form~(\ref{SS})
is obeyed with $\alpha=2$ and $z=4$ for $T_{F}=T_{c}$, for the 
quenches to $T_{F}<T_{c}$ the more general multiscaling form
\be
C({\vec k},t) \sim [L(k_{m}L)^{\frac{2-d}{d}}]^{\alpha(x)} F(x)
\label{MS}
\ee
is found, where $L(t) \sim t^{1/4}$, $k_{m}(t)$ is the peak wave
vector and $x=k/k_{m}$. The exponent $\alpha(x)$ is given by
\be
\alpha(x)= q + \varrho \varphi(x)
\label{alpha}
\ee
with

\be
q = \left \{
\begin{array}{cc}
2, &\mbox{for } 0 < T_F < T_c \\
0, &\mbox{for } T_F = 0
\end{array}
\right.
\ee

\be
\varrho = \left \{
\begin{array}{cc}
d-2, &\mbox{for } 0 < T_F < T_c \\
d, &\mbox{for } T_F = 0.
\end{array}
\right.
\ee
Furthermore, when $0 < T_F < T_c$ the function $\varphi(x)$ in~(\ref{alpha})
is given by
\be
\varphi(x) = \left \{
\begin{array}{cc}
\psi(x), &\mbox{for } x < x^*  \\
0, &\mbox{for } x > x^*  
\end{array}
\right.
\ee
with $x^*=\sqrt 2$ and
\be
\psi(x) = 1 - (1-x^2)^2
\label{psi}
\ee
while $\varphi(x) = \psi(x)$ for all values of $x$ when $T_F=0$.
Finally,

\be
F(x) = \left \{
\begin{array}{cc}
{T_F \over x^2}, &\mbox{for  } 0 < T_F < T_c \\
1, &\mbox{for  } T_F = 0.
\end{array}
\right.
\label{scf}
\ee

Leaving aside for the moment the apparent formal complication
of~(\ref{MS}- \ref{scf}), the important feature which is immediately
evident is that, contrary to~(\ref{eqalpha}), now there are three distinct
asymptotic behaviors for $T_F=T_c$, for $0<T_F<T_c$ and for $T_F=0$.
For $T_F=T_c$ the structure factor obeys standard scaling
with $\alpha = 2$ as in the NCOP case.
Instead for $T_F < T_c$ the exponent $\alpha$ 
depends on $x$ (Fig.~\ref{Fig1}) and
the scaling form~(\ref{MS}) involves two lengths, $k_{m}^{-1}(t)$
and $L(t)$, which differ by a logarithmic factor~\cite{Coniglio89} 

\be
(k_{m}L)^{4}=\log L^{d} + (2-d) \log (k_{m}L).
\label{log}
\ee
The functional form of $\alpha(x)$ is different for
$0 < T_F < T_c$ and for $T_F=0$.
This means that 
$T_F=T_c$ and $T_F=0$ are both unstable fixed points and
in between there is a new line of fixed points for $0< T_F < T_c$.
The temperature below the critical point is no more an
irrelevant variable.

If the $1/N$ expansion were a good systematic theory, the $1/N$ 
corrections ought to produce only minor quantitative changes
on the picture outlined above.
However, this expectation has not been fulfilled by the work of Bray and
Humayun (BH)~\cite{Bray92}, who found that for quenches to $T_F=0$  and COP 
standard scaling of
the form~(\ref{SS}) is restored in systems with any finite value
of $N$. The same result is very likely to apply
also to the quenches to $0<T_F<T_c$.
Therefore, the main qualitative features emerging in lowest order, like the
multiscaling behavior and the relevance of the temperature fluctuations,
are expected to be a peculiarity of the case with $N$ strictly infinite,
disappearing as soon as higher order corrections are taken into account.
In other words, the limits $N\to \infty$
and $t \to \infty$ do not commute for quenches to $T_F < T_c$.
In this paper we explore in some detail this phenomenon and we clarify
what dynamical process is really described when the $N\to \infty$ limit
is taken first. This helps to understand what correct use
is to be made of the large-$N$ model in this area of non equilibrium
statistical mechanics.

The gross features of what goes on in the quenches below $T_c$ 
can be described with the help of Figure 2. The phase-ordering process
of a system with finite $N$ is represented by path $I$ connecting the
disordered states $A$ to the ordered states $B$. These latter states are
mixtures of broken symmetry states. If, {\it after} equilibrium 
has been established, the $N \rightarrow \infty$ limit is taken 
as depicted by path $II$, the system is brought into a state $D$
which is the mixture of the $N \rightarrow \infty$ limits of each one
of the broken symmetry states. If, instead, the $N \rightarrow \infty$ limit
is taken {\it before} the quench, the process starts from a
disordered state $C$ of the system with infinitely many components
and the ensuing dynamical process does not
connect $C$ to $D$ along path $III$. As a matter of fact, the
process depicted by $III$ does not exist. Rather, the dynamical
evolution follows path $IV$ leading to low temperature states $E$
which are quite distinct from $D$. In other words, the system
with $N = \infty$ supports two different low temperature phases,
whose realization depends on the order of the limits
$t \rightarrow \infty$ and $N \rightarrow \infty$. The 
distinction between these two phases is reminiscent
of the difference between the zero field low temperature
states in the spherical model~\cite{Berlin52} and in 
the mean spherical model~\cite{Lewis52}.
In particular, states $E$ are very similar to the low temperature
states in the ideal Bose gas, as it will be clarified in the next section. 
The point to be stressed here is that in the static $1/N$
expansion states $A$ and $B$ are reached, respectively, from states $C$
and $D$, while $1/N$ corrections over states $E$ are not informative on
states $B$. This clarifies why the $1/N$ expansion
can be used for quenches to $T_{c}$, but not below $T_{c}$,
{\it as an approximation for processes with finite $N$}.

What, then, is the use of the  large-$N$ model for growth kinetics.
There is an obvious intrinsic interest, once it is clear that
although not describing a phase-ordering process of the usual type,
yet the model is well defined and describes the relaxation across
a phase transition. The growth process generated in the time
evolution can be studied in detail and produces non trivial 
behavior. The outcome is quite interesting
since, by modulating the initial noise and the final temperature,
remarkable crossover phenomena are obtained. Here is where the
model gives information also on systems with finite $N$, even if
it is not perturbatively close to the phase-ordering processes.
In fact, the phenomenology of the structure factor exhibits features
which are also found in the preasymptotic behavior of systems
with finite $N$~\cite{Castellano96a,Castellano96b}. Therefore,
through the large-$N$ model insight can be gained into the
very complex time regime preceding the onset of scaling in
realistic systems.

The paper is organized as follows.
In Section II the model is introduced, the non commutativity of the
$t \to \infty$, $N \to \infty$ limits is clarified and the nature of the
low temperature phases in the large-$N$ model is investigated.
In Section III the numerical solution for the structure factor is presented
and crossovers between different scaling behaviors are analyzed by means
of the multiscaling analysis.
Conclusions are presented in Section IV.

\section{The low temperature phases}
In the following we consider the relaxation dynamics of a system with an
$N$-component order parameter
${\vec \phi}({\vec x}) = (\phi_1({\vec x}),\ldots,\phi_N({\vec x}))$
which is initially prepared in a high temperature disordered state and
it is suddenly quenched to a lower temperature.
The evolution of the order parameter is governed by the time-dependent
Ginzburg-Landau model
\be
{\partial {\vec \phi}({\vec x},t) \over \partial t} =
-(i \nabla)^p 
{\partial {\cal H}[{\vec \phi},N] \over \partial {\vec \phi}({\vec x})}
+ {\vec \eta}({\vec x},t)
\label{eqofmotion}
\ee
where $p=0$ for NCOP, $p=2$ for COP,  and 
${\vec \eta({\vec x},t)}$ is the gaussian white
noise with expectations
\be
\left \{
\begin{array}{rcl}
<{\vec \eta}({\vec x},t)> & = & 0 \\
<\eta_{\alpha}({\vec x},t)\eta_{\beta}({\vec x}',t')> & =
& 2 T_F (i \nabla)^p\delta_{\alpha \beta} 
\delta({\vec x}-{\vec x}')\delta(t-t').
\end{array}
\right.
\ee
The free energy functional is of the form
\be
{\cal H}[{\vec \phi},N] = \int_V d^dx \left[
{1 \over 2} (\nabla {\vec \phi})^2 + {r \over 2} {\vec \phi}^2 +
{g \over 4N} ({\vec \phi}^2)^2
\right]
\label{freeenergy}
\ee
where $V$ is the volume of the system and $r<0$, $g>0$.
The order parameter probability distribution in the initial
state can be taken of the form
\be
P_0[{\vec \phi},N]= {1 \over Z_0} \exp{\left\{-{1 \over 2 \Delta}
\int d^dx {\vec \phi}^2({\vec x})\right\}}
\label{P0}
\ee
describing the absence of correlations at high temperature
\be
<\phi_{\alpha}({\vec x},0)\phi_{\beta}({\vec x}',0)> =
\Delta \delta_{\alpha \beta} \delta({\vec x}-{\vec x}').
\ee

As mentioned in the Introduction, if one wants to consider the
$N \to \infty$ limit, in order to determine the nature of the final
equilibrium state attention must be payed  to the order in which the
$N \to \infty$ and $t \to \infty$ limits are taken.
Let us consider first the sequence
$\lim_{N \to \infty}\lim_{t \to \infty}$.
Keeping $N$ finite, the equation of motion~(\ref{eqofmotion}) induces the
time evolution of the probability distribution from the
initial form~(\ref{P0}) toward the Gibbs state 
\be
P[{\vec \phi},t,N] \longrightarrow
P_{eq}[{\vec \phi},N]= {1 \over Z} \exp{\left\{-{1 \over T_F}
{\cal H}[{\vec \phi},N]\right\} }.
\label{Peq}
\ee
In the infinite volume limit
$P_{eq}[{\vec \phi},N]$ describes a disordered pure state
if $T_F$ is above $T_c$ and the $O(N)$ symmetrical 
mixture of the broken symmetry states if $T_F$ is below $T_c$.
If we now take the $N \to \infty$ limit (path of type 
$II$  in Fig.~\ref{Fig1}), for $T_F \geq T_c$ we obtain the pure phase $C$
\be
P_{eq}[{\vec \phi},\infty] =
{1 \over Z} \exp{ \left\{ -{1 \over 2 T_F} 
\sum_{\vec k}
(k^2 + r + gS) {\vec \phi}({\vec k}) \cdot {\vec \phi}(-{\vec k})
\right \} } 
\label{PeqC}
\ee
where $S$ is given by the self-consistency condition
\be
S= {1 \over V} \sum_{\vec k}
<\phi_{\beta}({\vec k}) \phi_{\beta}(-{\vec k})>
\ee
and $\beta$ denotes the generic component. 
Below $T_c$, denoting by $\vec{m}$ the expectation value of 
$\vec{\phi}(\vec{x})$ in the broken symmetry state, the
$N \rightarrow \infty$ limit $D$ of the mixture is obtained
\be
P_{eq}[{\vec \phi},\infty] = \int d \vec{m} \rho(\vec{m})
\mu [\vec{\phi} \mid \vec{m}]
\label{PeqD}
\ee
where $\rho(\vec{m})$ is the uniform probability density over
the sphere of radius $m$ and the pure state 
$\mu [\vec{\phi} \mid \vec{m}]$ is given by
\be
\mu [\vec{\phi} \mid \vec{m}] =
\delta [(\vec{\phi}-\vec{m}) \cdot \hat{m}]
{1 \over Z} \exp{ \left \{-{1 \over 2 T_F}
\sum_{\vec k}
(k^2 + r + gm^2 + gS_{\perp}) {\vec \phi}_{\perp}({\vec k}) \cdot
{\vec \phi}_{\perp}(-{\vec k})
\right \}}
\label{Peq2}
\ee
where $\vec{\phi}_{\perp}=\vec{\phi}-(\vec{\phi} \cdot \hat{m})\hat{m}$.
The quantities $S_{\perp}$ and $m$ are determined by the self-consistency
relations
\be
S_{\perp}= {1 \over V} \sum_{\vec k}<\phi_{\perp \beta}({\vec k})
\phi_{\perp \beta}(-{\vec k})>
\ee
\be
r + g(m^2+S_{\perp})=0.
\ee

In the end, computing averages with the weight 
functions ~(\ref{PeqC})~(\ref{PeqD}), we find the
well known result of the large-$N$ model
\be
<\phi_{\beta}({\vec k}) \phi_{\beta}(-{\vec k})> =
\left \{
\begin{array}{cc}
{T_F \over k^2 + r + gS}, &\mbox{for  } T_F \geq T_c \\
{T_F \over k^2} + m^2 \delta({\vec k}), &\mbox{for  } T_F < T_c \\
\end{array}
\right .
\label{N=infty}
\ee
for the equilibrium structure factor in states $C$ and $D$.
The average value of the order parameter is given by
\be
m^2 = -{r \over g} \left({T_c - T_F \over T_c} \right)
\ee
with
\be
T_c = - {r(d-2) \over g K_d \Lambda^{d-2}}
\label{T_c}
\ee
where $\Lambda$ is a wave vector cutoff and
$K_d = [2^{d-1} \pi^{d/2} \Gamma(d/2)]^{-1}$.
The $\delta({\vec k})$ term appearing in~(\ref{N=infty}), below $T_c$,
is the Bragg peak due to ordering in the low temperature phase $D$.

Let us now consider the limits in the opposite order.
Taking the $N \to \infty$ limit at the outset amounts to take the limit
on the equation of motion~(\ref{eqofmotion}) which becomes effectively
linearized
\be
{\partial \phi_{\beta}({\vec k},t) \over \partial t} =
-k^p 
[k^2+R(t)] \phi_{\beta}({\vec k},t)
+ {\vec \eta}_{\beta}({\vec k},t)
\label{eqlinearized}
\ee
with
\be
R(t)=r+gS(t)
\label{R}
\ee
and 
\be
S(t)={1 \over V} \sum_{\vec k}
<\phi_{\beta}({\vec k},t) \phi_{\beta}(-{\vec k},t)>.
\label{S}
\ee
Due to the linearity of Eq.~(\ref{eqlinearized}),
the time dependent probability distribution is
gaussian
\be
P[{\vec \phi},t,\infty] =
{1 \over Z(t)} \exp{\left\{-{1 \over 2}
\sum_{\vec k} C^{-1}({\vec k},t)
{\vec \phi}({\vec k}) \cdot {\vec \phi}(-{\vec k}) \right\}
}
\ee
with
\be
C({\vec k},t) =
<\phi_{\beta}({\vec k},t) \phi_{\beta}(-{\vec k},t)>.
\ee
Taking next the $t \to \infty$ limit 
\be
P[{\vec \phi},t,\infty] \rightarrow Q_{eq}[{\vec \phi}, \infty] = 
{1 \over Z} \exp{\left\{-{1 \over 2}
\sum_{\vec k} C^{-1}_{eq}({\vec k})
{\vec \phi}({\vec k}) \cdot {\vec \phi}(-{\vec k})\right\} }
\label{Qeq}
\ee
we obtain a gaussian state for any final temperature
and it is legitimate to ask whether still there is a phase
transition. 

From~(\ref{eqlinearized}) it is straightforward to 
obtain~\cite{Coniglio94} the equation of
motion for the structure factor
\be
{\partial C({\vec k},t) \over \partial t} =
-2 k^p  [k^2+R(t)] C({\vec k},t) + 2 k^p T_F
\label{eqmotion}
\ee
which, after equilibration is reached, yields
\be
C_{eq}({\vec k}) = {T_F \over k^2 + \xi^{-2}}
\ee
with NCOP and
\be
C_{eq}({\vec k}) =
\left \{
\begin{array}{cc}
C({\vec k}=0, t=0), &\mbox{for  } {\vec k} = 0 \\
{T_F \over k^2 + \xi^{-2}}, &\mbox{for  } {\vec k} \neq 0 
\end{array}
\right .
\ee
with COP, where $\lim_{t \to \infty} R(t)= \xi^{-2}$ is the inverse square
equilibrium correlation length.
From~(\ref{R}) 
\be
\xi^{-2} = r + {g T_F \over V} \sum_{{\vec k}} {1 \over k^2 + \xi^{-2}}
\label{xi2}
\ee
where the sum extends over all values of ${\vec k}$ for NCOP
and over ${\vec k} \neq 0$ for COP.
We now analyze Eq.~(\ref{xi2}) in the NCOP case referring to Appendix II
for the modifications in the argument required by the conservation law.
With $V$ finite the solution of Eq.~(\ref{xi2}) for $\xi$ is finite
for any temperature.
As expected, there is no phase transition in a finite volume system.
In the infinite volume limit (\ref{xi2}) becomes
\be
\xi^{-2} = r + g T_F B(\xi^{-2}) + {g T_F \over V \xi^{-2}}
\label{xi2bis}
\ee
where the function $B(x)$ defined by
\be
B(x)= \int {d^d k \over (2 \pi)^d } {1 \over k^2 + x}
\ee
is a monotonously decreasing function of $x$ with a maximum value
at $B(0) = K_d \Lambda^{d-2}/(d-2)$.
In writing~(\ref{xi2bis}) the ${\vec k}=0$ contribution to
the sum~(\ref{xi2}) has been explicitly separated out. Defining
$\gamma^2 = {T_F \over V \xi^{-2}}$ and introducing the temperature
$T_c = - {r \over gB(0)}$ which coincides with~(\ref{T_c}),
Eq.~(\ref{xi2bis}) can be rewritten as
\be
{1 \over g} \xi^{-2} = \left[
{r \over g} \left( {T_c - T_F \over T_c}\right) + \gamma^2
\right] + T_F \left[
B(\xi^{-2}) - B(0) \right]
\ee
with the solution (Appendix II)
\be
\left \{
\begin{array}{ccc}
\xi^{-2} > 0, & \gamma^2=0, &\mbox{for   }T_F > T_c \\
\xi^{-2} = 0, & \gamma^2=0, &\mbox{for   }T_F = T_c \\
\xi^{-2} = 0, & \gamma^2=-r/g \left(T_c -T_F \over T_c \right),
&\mbox{for   }T_F < T_c
\end{array}
\right .
\label{gamma}
\ee
which shows the existence of the phase transition at $T_c$.
For the structure factor this implies
\be
C_{eq}({\vec k}) = 
\left \{
\begin{array}{cc}
{T_F \over k^2 + \xi^{-2}} & T_F  \geq  T_c \\
{T_F \over k^2} + \gamma^2 \delta({\vec k}) & T_F < T_c 
\end{array}
\right .
\label{C}
\ee
and taking into account the form~(\ref{gamma}) of $\gamma^2$,
(\ref{C}) is identical to (\ref{N=infty}).
Thus, as far as the structure factor is concerned, the same result
is found irrespective of the order of the limits $t \to \infty$ and
$N \to \infty$.
However, comparing the states,
$Q_{eq}[\vec{\phi},\infty]$ coincides with $P_{eq}[\vec{\phi},\infty]$
above $T_c$, but not below where
\be
Q_{eq}[\vec{\phi},\infty]
= \frac{1}{\sqrt{2 \pi \gamma^{2} V}}
\exp \left \{-\frac{\phi^{2}(0)}
{2  \gamma^{2} V} \right \} \frac{1}{Z} \exp
\left \{ -\frac{1}{2} \sum_{\vec{k} \neq 0} C_{eq}^{-1}(\vec{k})
\vec{\phi}(\vec{k}) \cdot \vec{\phi}(-\vec{k}) \right \}.
\ee
This is a state exactly of the same form of the zero field
low temperature state in the mean spherical model, while state
$D$ is a mixture as in the spherical model~\cite{Kac77}.

Thus, despite the formal similarity, the Bragg peaks in~(\ref{N=infty})
and~(\ref{C}) have different physical meanings.
In the former case, it signals the formation of a mixture
of ordered states, while in the latter it is due to the macroscopic growth
of the ${\vec k}=0$ term in the sum~(\ref{xi2}).
What we have here is a low temperature phase obtained by condensation
of the fluctuations at ${\vec k}=0$, as in the ideal Bose gas,
with $C_{eq}({\vec k}=0)$ playing the role of the zero
momentum occupation number. 

Finally, notice that with $d=2$ the critical temperature
(\ref{T_c}) vanishes.
Hence, all states with $T_F>0$ are disordered states and the limits
$t \to \infty$ and $N \to \infty$ commute for the quenches to any finite
final temperature.
Conversely, $T_F = 0$ is not a critical temperature, rather it is an ordering
temperature.
Therefore, the quench to $T_F=0$ is an ordering process and the limits
are not supposed to commute in this case.

\section{Scaling behaviors}
In this section we investigate the time evolution of the structure factor
by solving numerically the equation of
motion~(\ref{eqmotion}). It is convenient to comment beforehand on the
structure of this solution. Integrating~(\ref{eqmotion}) the
structure factor can be written as the sum of two contributions
\be
C(\vec{k},t) = C_{0}(\vec{k},t) + C_{T}(\vec{k},t)
\label{splitting}
\ee
with
\be
C_{0}({\vec k},t)  =  \Delta \exp \left \{ -2 k^p \int_0^t ds \left[
k^2+R(s)\right] \right \} 
\label{czero}
\ee
and
\be
C_{T}({\vec k},t)  =  2 T_F k^p \int_0^t dt^{\prime}
\exp \left\{ -2 k^p \int_{t^{\prime}}^t ds \left[ k^2+R(s)\right]
\right \}.
\label{CT}
\ee
The asymptotic behavior is analytically accessible and it has been
derived in detail in Ref.~\cite{Coniglio94}. For quenches to
$T_F \le T_c$ with NCOP, the large time behavior is given by
\be
C_{0}({\vec k},t) = \Delta L^{\omega} f_{0}(x)
\ee
\be
C_{T}({\vec k},t) = T_F L^2 f_T(x)
\ee
with $L(t) = t^{1/2}$, $x=kL$ and
\be
\omega = \left \{
\begin{array}{cc}
4-d, &\mbox{for } T_F = T_c \\
d, &\mbox{for } T_F < T_c
\end{array}
\right.
\ee
\be
f_{0}(x)= e^{-x^{2}}
\ee
\be
f_{T}(x) = \int_{0}^{1} dy (1-y)^{-\omega/2} e^{-x^{2}y}.
\ee
Notice that both contributions are in the scaling form~(\ref{SS})
with $C_{T}({\vec k},t)$ dominating in the quenches to $T_c$
while $C_{0}({\vec k},t)$ dominates for $T_F < T_c$. One may go
one step further regarding the order parameter as the sum of two
contributions $\vec{\phi} = \vec{\sigma} + \vec{\zeta}$ whose
correlations account for the two pieces in the structure factor
$C_{0}({\vec k},t) = <\sigma_{\beta} \sigma_{\beta}>$,
$C_{T}({\vec k},t) = <\zeta_{\beta} \zeta_{\beta}>$
and $ <\sigma_{\beta} \zeta_{\gamma}> = 0$. Therefore,
for $T_F < T_c$ a sort of two fluids picture of the quench is
obtained, with the condensate $\vec{\sigma}$ and the thermal
fluctuations $\vec{\zeta}$ having a distinct individuality due to
the different scaling properties. Notice that the irrelevance
of the thermal fluctuations is due to $\vec{\sigma}$ dominating
$\vec{\zeta}$ and obeying the zero temperature equation of
motion.

With COP there is additional structure since the thermal
contribution contains itself two different pieces
\be
C_{T}({\vec k},t) = \left \{
\begin{array}{cc}
C_{<}({\vec k},t), &\mbox{for } k < x^* k_m(t) \\
C_{>}({\vec k},t), &\mbox{for } k > x^* k_m(t)
\end{array}
\right.
\ee
with the asymptotic behaviors
\be
C_{0}({\vec k},t) = \Delta L^{\rho \psi(x)}
\label{011}
\ee
\be
C_{<}({\vec k},t) = \frac {T_F}{x^{2}} L^{2+\rho \psi(x)}
\label{012}
\ee
\be
C_{>}({\vec k},t) = \frac {T_F}{x^{2}} L^{2}
\label{013}
\ee
\be
\rho = \left \{
\begin{array}{cc}
0, &\mbox{for } T_F = T_c \\
d-2, &\mbox{for } 0 < T_F < T_c \\
d, &\mbox{for } T_F =0.
\end{array}
\right.
\label{014}
\ee
For simplicity, in writing~(\ref{011}-\ref{013}) we have neglected
the logarithmic difference between $L(t)$ and $k_{m}(t)$, putting 
$x=kL$ and $L(t) = t^{1/4}$. The novelty is in the
region $0 < T_F < T_c$, where there is a sharp distinction between
what happens for $x < x^*$ and for $x > x^*$. In the first case 
$C_{<}({\vec k},t)$ dominates over $C_{0}({\vec k},t)$, while in
the second case $C_{>}({\vec k},t)$ dominates over $C_{0}({\vec k},t)$.
Therefore, with COP we are led to regard the order parameter as
the sum of three contributions $\vec{\phi} =\vec{\sigma}+ \vec{\zeta_{<}} +
\vec{\zeta_{>}}$ which, again, are characterized by distinct scaling 
properties and whose correlations are responsible, respectively, for
$C_{0}({\vec k},t)$, $C_{<}({\vec k},t)$ and $C_{>}({\vec k},t)$.
The remarkable qualitative difference with the NCOP case, is 
that now the condensate also is of thermal origin because the Bragg
peak is formed by $\vec{\zeta_{<}}$. Hence, the temperature is not
an irrelevant variable. Furthermore, the thermal fluctuations 
$\vec{\zeta_{>}}$ which obey standard scaling, as the time goes
on propagate from the large wave vectors toward the small wave
vectors, following a pattern which is important,
as we shall see below, for understanding 
what happens in the realistic systems. Finally, the 
$\vec{\sigma}$ contribution which was responsible for the condensate
with NCOP, here is subdominant, but it can give rise to 
interesting preasymptotic behaviors if $\Delta$ and $T_F$ are
appropriately chosen. 

In the numerical study of the structure factor, we shall 
devote particular attention to the transition from preasymptotic 
to asymptotic features.
This we do both for NCOP and COP with $d=3$ and $d=2$.
Parameters of the quench are the final temperature $T_F$ and the strength
$\Delta$ of the fluctuations in the initial state~(\ref{P0}). In particular,
we will consider the two cases $\Delta = 0$ (small $\Delta$) and
$\Delta  = -r/g$ (large $\Delta$), where $\sqrt{-r/g}$ is the
equilibrium value of the order parameter at zero temperature.
The final temperature of the quench is important 
in two  respects. First of all, for $T_F > T_c$
the correlation length $\xi$ is finite, while for $T_F \leq T_c$ 
it is infinite. This is important because
the general structure of the time
evolution is determined by the relation between a microscopic length
$L_0$ and the correlation length $\xi$ in the final equilibrium state.
The initial fast transient, with no scaling, lasts up to some
time $t_0$.
At this point equilibrium is established over the length scale $L_0$ 
and, if this is of the order of magnitude of $\xi$, final
equilibrium is reached over the whole system as well.
Instead, if $\xi \gg L_0$, a second regime is entered~\cite{Mazenko82}, 
during which the scaling relations (1) and (2) hold. This
lasts up to the time $t_1$ such that
$L(t_1) \simeq \xi$, when global equilibrium is again established.
Clearly, if $\xi$ is infinite, equilibrium is never reached and the
scaling regime lasts forever. The second important feature involving 
the final temperature is that, while for $T_F \geq T_c$ only the
thermal fluctuations grow, for $T_F < T_c$ there is growth of
both the condensate and thermal fluctuations.

\subsection{Multiscaling analysis}

The interplay of all these elements produces a rich variety of
behaviors which can be efficiently monitored through 
the multiscaling analysis.
This works as follows:
let us assume that the structure factor can be written in the general
multiscaling form
\be
C({\vec k},t) = [{\cal L}_{1}(t)]^{\alpha(x)} F(x)
\label{MS2}
\ee
with $x=k{\cal L}_{2}(t)$ and where ${\cal L}_{1}(t)$
and ${\cal L}_{2}(t)$ are two lengths. The functions $\alpha(x)$
and $F(x)$ are to be determined.
In order to check on the assumption, the time axis is divided in intervals
$(t_i,t_i+\tau_i)$ which in practice may also be of variable length
$\tau_i$, and within each interval the logarithm of $C({\vec k},t)$
is plotted against the logarithm of ${\cal L}_{1}(t)$ for a fixed value of
$x$.
By measuring the slope and the intercept of the plot $\alpha(x,t_i)$ and
$F(x,t_i)$ are obtained.
The procedure is then repeated for different values of $x$ and over
different time intervals.
If $\alpha(x,t_i)$ and $F(x,t_i)$ do not depend on $t_i$ the
assumption~(\ref{MS2}) is correct and scaling holds. Specifically,
standard scaling is the case where ${\cal L}_{1}(t)={\cal L}_{2}(t)$
and $\alpha(x)$ does not depend on $x$.
In the case of multiscaling $\alpha(x)$ does depend on $x$
and the two lengths ${\cal L}_{1}(t)$ and ${\cal L}_{2}(t)$
differ by a logarithmic factor.
Conversely, if $\alpha(x,t_i)$ and $F(x,t_i)$ do depend on $t_i$,
scaling does not hold.
Notice that in the case that equilibrium
has been reached, the disappearance of the time dependence shows up
as $\alpha(x,t_i) \equiv 0$.
In the following we shall not be interested in the determination
of $F(x)$ and we shall concentrate on $\alpha(x)$.

As an example, consider the quench to $T_F=0$.
With NCOP the exact form of the structure factor 
is given by~\cite{Coniglio89}
\be
C({\vec k},t) = \Delta \exp{\left\{-[Q(t)+(kL)^2]\right\}}
\ee
where $Q(t)$ is a function of time and $L(t) = t^{1/2}$.
Hence, we have the natural choice ${\cal L}_{2}(t) = L(t)$ and,
using~(\ref{MS2}),
\be
\alpha(x,t)= - {Q(t) \over \log {\cal L}_{1}(t) }
\label{alphancop}
\ee
which shows that if there is scaling it is of the standard
type. This occurs for large time where 
$Q(t) = -d \log L(t)$ suggests to take ${\cal L}_{1}(t) =
{\cal L}_{2}(t) = L(t)$, yielding $\alpha(x,t)=d$.
Conversely, for short time, if $\Delta$ is small enough to allow 
for the application of the linear approximation,
we have  $Q(t)=2rt$ and there is no scaling since the time dependence
does not drop out
\be
\alpha(x,t) = -{2rt \over \log L(t)}.
\label{3.7}
\ee
With COP, instead, the exact form of the structure factor is given 
by~\cite{Coniglio89}
\be
C({\vec k},t) = \Delta \exp{(k_mL)^4 \psi(k/k_m)}
\ee
where $k_m(t)$ is the peak wave vector, $L(t)=t^{1/4}$
and $\psi$ is given by~(\ref{psi}).
Choosing ${\cal L}_{2}(t)= k_{m}^{-1}(t)$ we find
\be
\alpha(x,t)= { \left[ k_m(t)L(t) \right]^4 \psi(x)
\over \log {\cal L}_{1}(t) }
\label{3.9}
\ee
which shows that the $x$ dependence cannot be eliminated and therefore
that scaling can only be of the multiscaling type.
This occurs for large time 
with ${\cal L}_{1}(t) = \left(k_m^{2-d} L^2\right)^{1/d}$ yielding
\be
\alpha(x,t)=d \psi(x).
\label{3.10}
\ee
Using~(\ref{log}), the two lengths ${\cal L}_{1}(t)$ and
${\cal L}_{2}(t)$ are related by ${{\cal L}_{1}(t) \over
{\cal L}_{2}(t)} \sim (\log L)^{2/d}$.
Conversely, in the very early stage where the linear approximation
holds ${\cal L}_{2}(t)= k_{m}^{-1} =\sqrt{-2/r}$ is time independent
and ${\cal L}_{1} \sim t^{1/2d}$ yielding
\be
\alpha(x,t) \sim  {t \over \log t} \psi(x)
\label{linearalpha}
\ee
which displays the absence of scaling through a time dependent prefactor
in front of $\psi(x)$.

\subsection{Evolution of $\alpha(x,t)$}

In the following we illustrate the evolution of $\alpha(x,t)$ obtained
numerically over a sequence of time intervals, with $r=-1$ and $g=1$.

\bigskip

{\bf NCOP} $d=3$

\bigskip

\noindent Let us begin with NCOP in the three dimensions. 
The critical temperature is finite $T_c=2\pi$,  
the exponent $\alpha$ is obtained setting
${\cal L}_{1}(t)={\cal L}_{2}(t)$ and extracting this length from 
the inverse of the halfwidth of the structure factor.
According to the general outline presented at the beginning
of this section, we expect to detect the establishment of equilibrium
for $T_F>T_c$ through the vanishing of $\alpha$ and the scaling
behavior lasting indefinitely 
for $T_F\leq T_c$, through the disappearance of the time dependence
around a non vanishing value of $\alpha$.
This is clearly illustrated in Fig.~\ref{Fig3}, where $\alpha$ is plotted
for a quench well above $T_c$ with $T_F=20$ and for a quench to $T_c$.
In the first case the lines rapidly collapse on $\alpha=0$, while in the
second case the collapse is on $\alpha=2$.
It is then interesting to consider the case of a final temperature slightly
above $T_c$, which corresponds to a large but finite $\xi$.
Also in this case $\alpha$ is expected to collapse eventually
on $\alpha=0$, however as long as $L(t)$ is large but smaller than $\xi$,
one expects to observe a behavior similar to the one in the quench to
$T_c$.
Indeed, this is what happens in Fig.~\ref{Fig4} (panel $a$), 
obtained by plotting $\alpha$ for $T_F=6.35$.
After the initial transient there is a collapse on $\alpha=2$,
revealing critical scaling.
However, this does not last indefinitely as in the case of the quench
to $T_c$, but it lasts for the time necessary for $L(t)$ to catch up
with $\xi$.
After this a new transient sets in and the eventual collapse
on the equilibrium value $\alpha = 0$ takes place.
For $T_F<T_c$ the behavior of $\alpha$ is quite similar to the
one for $T_F=T_c$, since in both cases $\xi=\infty$.
The only difference is that the asymptotic
behavior produces collapse on the value $\alpha = d =3$.
For $T_F$ slightly below $T_c$, e.g $T_F=6.20$ in Fig.~\ref{Fig4}
(panel $b$),
there is crossover from critical scaling with $\alpha=2$ to the final
value with $\alpha=3$. In general,
the asymptotic behavior for $0<T_F<T_c$ and $T_F=0$ 
is the same (Fig.~\ref{Fig5}),
confirming the irrelevance of thermal fluctuations. In all 
quenches considered, the variation of the size $\Delta$ of initial
fluctuations does not produce significant differences.

\bigskip

{\bf NCOP} $d=2$

\bigskip

\noindent In two dimensions the critical temperature~(\ref{T_c}) vanishes.
So, for any $T_F>0$ one should observe a behavior for $\alpha$
similar to the one obtained with $d=3$ and $T_F>T_c$.
This is the case for quenches with a final temperature well above
zero, where behaviors of $\alpha$ very close to the one in 
panel $a$ of Fig.~\ref{Fig3}
are obtained. Similarly, for the quench to $T_F=0$ the same behavior 
of Fig.~\ref{Fig5} (panel $b$) is obtained,
except that now the lines collapse on $\alpha=d=2$.

A case to be considered separately is when $T_F$ is finite but
very close to zero.
Fig.~\ref{Fig6} corresponding to $T_F=0.3$ and $T_F=10^{-6}$, 
displays an intermediate
scaling behavior with $\alpha=2$ preceding the eventual collapse on
the equilibrium value $\alpha=0$.
At first sight this looks like the behavior of Fig.~\ref{Fig4} (panel $a$)
in the quench to a final temperature slightly above $T_c$.
However, the interpretation is more subtle, since what is growing here,
as long as $L(t)<\xi$, are not the critical fluctuations but the
condensate.
The distinction between condensation and critical behavior is actually
impossible to make with NCOP on the basis of the value of $\alpha$, since
with $d=2$ in both cases $\alpha=2$.
This remark will become clear with COP, because in that case the growth of the
critical fluctuations is associated to standard scaling, while the growth
of the condensate gives rise to multiscaling.

\bigskip

{\bf COP} $d=3$

\bigskip

\noindent We now move on to the case of COP with both small 
and large initial fluctuations.
As discussed above, $\alpha(x,t)$ is extracted by defining $x=k/k_m$,
where $k_m(t)$ is the peak wave vector, and plotting $\log C$
vs. $\log {\cal L}_{1}(t)$ with
${\cal L}_{1}(t) = (k_m^{2-d}L^2)^{1/d}$ and $L(t)=t^{1/4}$.
For quenches to $T_F \geq T_c$ the behavior of $\alpha(x,t)$ essentially
follows the same pattern as in the NCOP case.
Apart from some differences in the time dependent transients, 
again for $T_F > T_c$ and for
$T_F=T_c$ curves collapse, respectively,
on $\alpha=0$ and $\alpha=2$. In Fig.~\ref{Fig7} we illustrate what 
happens in the quench slightly above (panel $a$) and slightly below 
(panel $b$) $T_c$. In both cases there is a preasymptotic
standard scaling behavior with $\alpha=2$ due to the
critical point in the neighborhood, followed by the incipient
crossover toward the asymptotic behavior. Above $T_c$ this is
$\alpha =0$, as revealed by $\alpha$ deviating downward, while
below $T_c$ the deviation occurs upwardly, for $x < x^*$, toward
the asymptotic form of Fig.~\ref{Fig1}.

Where the difference between NCOP and COP becomes remarkably
evident is in the quenches well below $T_c$. 
Let us first consider (Fig.~\ref{Fig8}) $T_F=0$. 
After a time dependent transient (panel $a$), 
which for small $\Delta$ in the early
stage is well described by (\ref{linearalpha}),
the curves of $\alpha(x,t)$ collapse (panel $b$) on the 
limiting curve~(\ref{3.10}) depicted in Fig.~\ref{Fig1}.
Instead, for $T_F=1$ (Fig.~\ref{Fig9}), the collapse occurs
on the finite temperature asymptotic form of Fig.~\ref{Fig1}, 
with minor differences
in the transient due to the size of $\Delta$. However,
going to a temperature much lower but finite (Fig.~\ref{Fig10}),
while for $\Delta =0$ (panel $a$) $\alpha$ follows the
same pattern as in Fig.~\ref{Fig9}, the behavior
in panel $b$ with $\Delta =1$ is drastically different.
What we have here is multiscaling as in the quench to $T_F=0$,
for $x<x^*$, and standard scaling with $\alpha=2$ for
$x>x^*$. All these features can be accounted for on the
basis of the discussion made at the beginning of this section.
As long as $T_F$ is sufficiently large or $\Delta$ small,
as in Fig.~\ref{Fig9} and in panel $a$ of Fig.~\ref{Fig10},
only $C_{<}(\vec{k},t)$ and $C_{>}(\vec{k},t)$ contribute
to $\alpha(x,t)$ producing the characteristic behavior
of Fig.~\ref{Fig1}. However, when $\Delta$ is finite and
$T_F$ small enough, there can be a sizable interval of
time during which $C_{0}(\vec{k},t)$ dominates over
$C_{<}(\vec{k},t)$, for $x<x^*$, producing the scaling pattern
of panel $b$ in Fig.~\ref{Fig10}. This behavior is preasymptotic
and eventually the crossover to the pattern of 
Fig.~\ref{Fig9} takes place, as $C_{<}(\vec{k},t)$ grows
large enough to overtake $C_{0}(\vec{k},t)$. In our
numerical solution the computation was not run long enough to actually
detect this crossover, but 
it is clear that by modulating the parameters of the quench, $\Delta$
and $T_F$, the crossover time can be varied at will.

\bigskip

{\bf COP} $d=2$

\bigskip

\noindent It is now interesting to see how this 
variety of behaviors is affected
by pushing the critical temperature to zero in two dimensions.
The novelty with respect to the previous case is that now the line of fixed
points in between $T_F=0$ and $T_c$ has disappeared and with it, supposedly,
also the associated asymptotic behavior.
Actually, only the fixed point at $T_F=0$ has survived.
Thus we should observe either the relaxation to equilibrium for $T_F>0$,
or the multiscaling behavior for $T_F=0$.
Indeed, for temperatures like $T_F=10$ the collapse on 
$\alpha=0$ is observed.
Similarly, the quench to $T_F=0$ produces a behavior
identical to the one in Fig.~\ref{Fig8}, except that the peak 
value of $\alpha$ is given by 2 in place of 3.
The interesting novelties arise when quenches to small, but finite, $T_F$ are
considered. In fact, when $T_F$ is low $\xi$ is large and 
an intermediate scaling regime preceding the final equilibration is expected.
The question is, what it will be like.
For $T_F=10^{-6}$ (Fig.~\ref{Fig11}) and $\Delta=0$ there is standard scaling
with $\alpha=2$, while for $\Delta=1$ the pattern is identical
to the one in panel $b$ of Fig.~\ref{Fig10}, except for the obvious
modification $\alpha(x=1)=d=2$. The first observation is that
these are preasymptotic behaviors, since eventually $\alpha$ 
must vanish. The second is that this phenomenology can be understood
regarding, as long as $L(t)< \xi$, the structure factor as made up of
the three contributions (\ref{011}-\ref{013}). Namely,
as long as $L(t)< \xi$, there is growth of the condensate and
of thermal fluctuations, as in the quench below the critical
point. Thermal fluctuations are clearly due to $T_F>0$, while
the growth of the condensate originates from the underlying
fixed point being at $T_F=0$. The net result is that the intermediate
scaling behavior with $\Delta=0$ is due to $C_{<}(\vec{k},t)$
and $C_{>}(\vec{k},t)$, which scale both like $L^{2}$ since
$\rho$ vanishes when $d=2$. Instead, with $\Delta=1$ we are
confronted again with a situation where, in the time of the 
computation, $C_{0}(\vec{k},t)$ dominates over $C_{<}(\vec{k},t)$
producing the pattern of panel $b$. As a matter of fact this is
a pre-preasymptotic scaling behavior, since when $C_{<}(\vec{k},t)$
overtakes $C_{0}(\vec{k},t)$ a behavior of the type in panel $a$
is expected to occur before the eventual relaxation to a vanishing
$\alpha$.

\section{Conclusions}

In this paper we have investigated the origin of the non commutativity
of the limits $t \rightarrow \infty$ and $N \rightarrow \infty$
in the dynamics of the first order transitions. The main result
is that when the $N \rightarrow \infty$ limit is taken first
the underlying phase transition, which we have called condensation,
is qualitatively different from the usual process of ordering
obtained with the limits in the opposite order.
In particular, condensation in conjunction with COP gives rise,
for reasons which are not yet clear, to two phenomena which are
strikingly different from  what one has in phase-ordering,
namely i) multiscaling and
ii) relevance of the thermal fluctuations.  We have then
proceeded to an extensive investigation of the scaling properties of the
structure factor in the large-$N$ model for quenches to a final
temperature greater, equal or lower than $T_c$. We have found a rich variety
of behaviors, which can be studied in great detail through the
multiscaling analysis. Particularly interesting is the existence
of preasymptotic scaling which can be explained through the 
competition between different components of the order parameter
with distinct scaling properties.

Even though it is quite clear that the large-$N$ model is not
perturbatively close to the phase-ordering processes in realistic
systems, in concluding the paper we wish to elaborate on the
connections that nonetheless exist. In order to do this we use
the BH model~\cite{Bray92} as an intermediate step. As mentioned above, in
this model the structure factor for the quenches to $T_F=0$ with
COP displays standard scaling for any finite value of $N$. 
The discussion in the previous section on the behavior of $\alpha$
in the quenches to $T_F=10^{-6}$ helps to understand how this
comes about. By integrating formally the equation of motion,
the BH structure factor can be written as the sum of two 
contributions
\be
C(\vec{k},t) = C_{0}(\vec{k},t) + C_{N}(\vec{k},t)
\label{41}
\ee
where $C_{0}(\vec{k},t)$ is given by (\ref{czero}) while 
\be
C_{N}(\vec{k},t)  = 
-{2 \over N} k^p \int_{0}^{t} dt^{\prime} R(t^{\prime})
{C_{0}(\vec{k},t) \over C_{0}(\vec{k},t^{\prime})}
D(\vec{k},t^{\prime})
\label{BH}
\ee
with
\be
D(\vec{k},t)=
\int {d^dk_1 \over (2 \pi)^d}
{d^dk_2 \over (2 \pi)^d} C({\vec k}-{\vec k}_1,t)
C({\vec k}_1-{\vec k}_2,t) C({\vec k}_2,t)
\ee
contains the nonlinearity and $R(t)$ is given by (\ref{R}) and (\ref{S}).
Although Eq. (\ref{splitting}) refers to a quench to $T_F > 0$
with $N = \infty$ and Eq. (\ref{41}) to a quench to $T_F=0$
and $N$ finite, the mechanism regulating the competiton 
between the two terms is the same.
In particular, in both cases $C_{0}(\vec{k},t)$ can compete
with the second term only for $x<x^*$. BH have shown 
that $C_{N}(\vec{k},t)$ asymptotically obeys standard scaling
with $\alpha=d$. It is then clear that by choosing $\Delta$
and $N$ properly there may be a preasymptotic regime during which 
$C_{0}(\vec{k},t)$ dominates for $x<x^*$, much in the same way
as in the previous section $C_{0}(\vec{k},t)$ was found to 
dominate over $C_{T}(\vec{k},t)$. Here $1/N$ plays a role similar
to that of $T_F$ and the crossover time depends on both $\Delta$ and
$N$~\cite{Bray92,Castellano96a}. With $N=100$ and $d=2$ 
(Fig.~\ref{Fig12}) a behavior for $\alpha(x,t)$ is obtained which
is practically the same of that in Fig.~\ref{Fig11}. In order
to complete the picture, we reproduce in Fig.~\ref{Fig13} the
behavior of $\alpha(x,t)$ for the scalar system obtained in
Ref.~\cite{Castellano96b} by the simulation of Eq. (\ref{eqofmotion})
with $N=1$ and $d=2$. Again the same pattern is found, revealing
that the same mechanism is operating also in this case. Therefore,
one may conclude that the behavior of the large-$N$ model
associated to $C_{0}(\vec{k},t)$ describes what happens in the
preasymptotic regime also in the phase-ordering processes
over the shrinking range of wave vectors with $k < x^* k_{m}(t)$.
In other words, the asymptotic regime is preceeded by a time regime
where phase-ordering over the short length scale seems to coexist
with condensation over the large length scale.
This is not surprising for NCOP since correlations are established over
regions of size $L(t)$, and the statistics can be expected to become gaussian
over distances larger than $L(t)$. For COP it is less straightforward,
although the occurrence of gaussian statistics on large length scales can
be detected much more easily through the appearence of multiscaling
behavior.

\section{Appendix I}
From~(\ref{SS}) the real space scaling form of the correlation function
is given by
\be
G({\vec r},t) = r^{\alpha-d} g(r/L(t))
\label{SSReal}
\ee
where $g(x)$ is the scaling function and $g(0)$ is a finite quantity.
For $r\ll L(t)$ (\ref{SSReal}) gives the equilibrium decay of the correlation
function
\be
G_{eq}({\vec r}) \sim r^{\alpha-d}.
\label{Geq}
\ee
On the other hand, the correlation function on a fractal~\cite{Vicsek89},
decays as
\be
G_{eq}({\vec r}) \sim r^{2(D-d)}
\label{Geqfr}
\ee
where $D$ is the fractal dimensionality. Comparing (\ref{Geqfr}) with
(\ref{Geq}) the fractal dimensionality of the correlated regions is given by
$D={1 \over 2}(\alpha + d)$ and using~(\ref{eqalpha})
\be
D = \left \{
\begin{array}{cc}
{1 \over 2}(2 + d -\eta), &\mbox{for } T_F = T_c \\
d, &\mbox{for } T_F < T_c.
\end{array}
\right.
\ee

\section{Appendix II}

{\bf NCOP}

\bigskip

\noindent Equation~(\ref{xi2bis}) can be rewritten as
\be
\xi^{-2} = r + g T_F B(0) + {g T_F \over V \xi^{-2}}
+g T_F \left[B(\xi^{-2})-B(0) \right]
\label{xi2ter}
\ee
and using $r+gT_cB(0)=0$
\be
{\xi^{-2} \over g} = c + T_F \left[B(\xi^{-2})-B(0) \right] 
\label{xi2quater}
\ee
where
\be
c=-{r \over g} \left({ T_F -T_c \over T_c} \right) + {T_F \over V \xi^{-2}}.
\label{c}
\ee
Since $\left[B(\xi^{-2})-B(0) \right]$ is a non 
positive monotonously decreasing function,
there is a positive solution of~(\ref{xi2quater}) for $c>0$, 
a vanishing solution for $c=0$ and no solution for $c<0$. 
For $T_F < T_c$ the quantity $c$ cannot be positive, because
in that case $\xi^{-2}$ would be positive and the second term 
in the right hand side of (\ref{c})
would vanish in the infinite volume limit, producing $c<0$.
Therefore, $c$ can only vanish implying
\be
\lim_{V \to \infty} {T_F \over V \xi^{-2}} = \gamma^2 = -
{r \over g} \left({ T_c -T_F \over T_c} \right).
\label{limV}
\ee

{\bf COP}

\bigskip

\noindent With conserved order parameter, Eq.~(\ref{xi2}) is replaced by
\be
\xi^{-2} = r + {g T_F \over V} \sum_{{\vec k}\neq 0} {1 \over k^2 + \xi^{-2}}
\label{xi22}
\ee
which allows for a solution $\xi^{-2} > - k^2_{min}$, where $k_{min} \sim V^{-1/d}$
is the minimum value of the wave vector. For
\be
T_F < {\tilde T}_c = -{r \over g}  \left[ {1\over V} \sum_{{\vec k}\neq 0} {1 \over k^2}
\right]^{-1}
\label{Ttilde}
\ee
the solution is negative $-k^2_{min} < \xi^{-2} < 0$.
In the infinite volume limit ${\tilde T}_c \to T_c$ and
$\xi^{-2} \to -k^2_{min}$. Thus, rewriting (\ref{xi22}) as 
\be
\xi^{-2} = r + g T_F B(\xi^{-2}) + {g T_F \over V (k^2_{min} +\xi^{-2})}
\ee
we can analyze this equation exactly as in the NCOP case, obtaining for $T_F < T_c$ the 
analog of (\ref{limV})
\be
\lim_{V \to \infty} {T_F \over V (k^2_{min} + \xi^{-2})} = \gamma^2 = -
{r \over g} \left({ T_c -T_F \over T_c} \right).
\ee

%\clearpage
\bc
\bf FIGURE CAPTIONS
\ec

\begin{enumerate}
\item \label{Fig1}
Spectrum of the multiscaling exponent $\alpha(x)$ for quenches 
with COP and $d=3$.

\item \label{Fig2}
Schematic representation of the relaxation processes in the systems
with $N$ finite and $N = \infty$.

\item \label{Fig3}
Evolution of $\alpha(x)$ for NCOP, $d=3$. ($a$) $T_F \gg T_c$ 
($b$) $T_F=T_c$.  Different curves refer to a sequence of time intervals
growing exponentially with the label.
In this and all other figures except Fig.~\ref{Fig8}, very early times are not
shown for simplicity.

\item \label{Fig4}
Evolution of $\alpha(x)$ for NCOP, $d=3$ for quenches to $T_F$ slightly
above and slightly below $T_c$.
Symbols and time intervals are related as in Fig.~\ref{Fig3}.

\item \label{Fig5}
Evolution of $\alpha(x)$ for NCOP, $d=3$ for quenches to $0<T_F<T_c$ 
and $T_F=0$.
Symbols and time intervals are related as in Fig.~\ref{Fig3}.

\item \label{Fig6}
Evolution of $\alpha(x)$ for NCOP, $d=2$ for quenches to $T_F = 0.3$ 
and $T_F=10^{-6}$.
Symbols and time intervals are related as in Fig.~\ref{Fig3}.

\item \label{Fig7}
Evolution of $\alpha(x)$ for COP, $d=3$ for quenches to $T_F$  slightly
above and slightly below $T_c$.
Symbols and time intervals are related as in Fig.~\ref{Fig3}.

\item \label{Fig8}
Evolution of $\alpha(x)$ for COP, $d=3$ for a quench to $T_F=0$.
($a$) early times ($b$) intermediate to late times.
Different curves refer to a sequence of time intervals
growing exponentially with the label.

\item \label{Fig9}
Evolution of $\alpha(x)$ for COP, $d=3$, $T_F=1$.
($a$) zero initial fluctuations, ($b$) large  initial fluctuations.
Symbols and time intervals are related as in Fig.~\ref{Fig3}.

\item \label{Fig10}
Evolution of $\alpha(x)$ for COP, $d=3$, $T_F=10^{-6}$.
($a$) zero initial fluctuations, ($b$) large  initial fluctuations.
Symbols and time intervals are related as in Fig.~\ref{Fig3}.

\item \label{Fig11}
Evolution of $\alpha(x)$ for COP, $d=2$, $T_F=10^{-6}$.
($a$) zero initial fluctuations, ($b$) large  initial fluctuations.
Symbols and time intervals are related as in Fig.~\ref{Fig3}.

\item \label{Fig12}
Evolution of $\alpha(x)$ for the solution of the Bray-Humayun model
with $N=100$, $d=2$, $T_F=0$.
($a$) small initial fluctuations, ($b$) large  initial fluctuations.
Different curves refer to a sequence of time intervals
growing exponentially with the label.

\item \label{Fig13}
Evolution of $\alpha(x)$ obtained in Ref.~\cite{Castellano96a}
by simulation of a system with $N=1$, $d=2$ and $T_F=0$.
($a$) small initial fluctuations, ($b$) large  initial fluctuations.
Different curves refer to a sequence of time intervals
growing exponentially with the label.

\end{enumerate}

\clearpage

%\end{multicols}
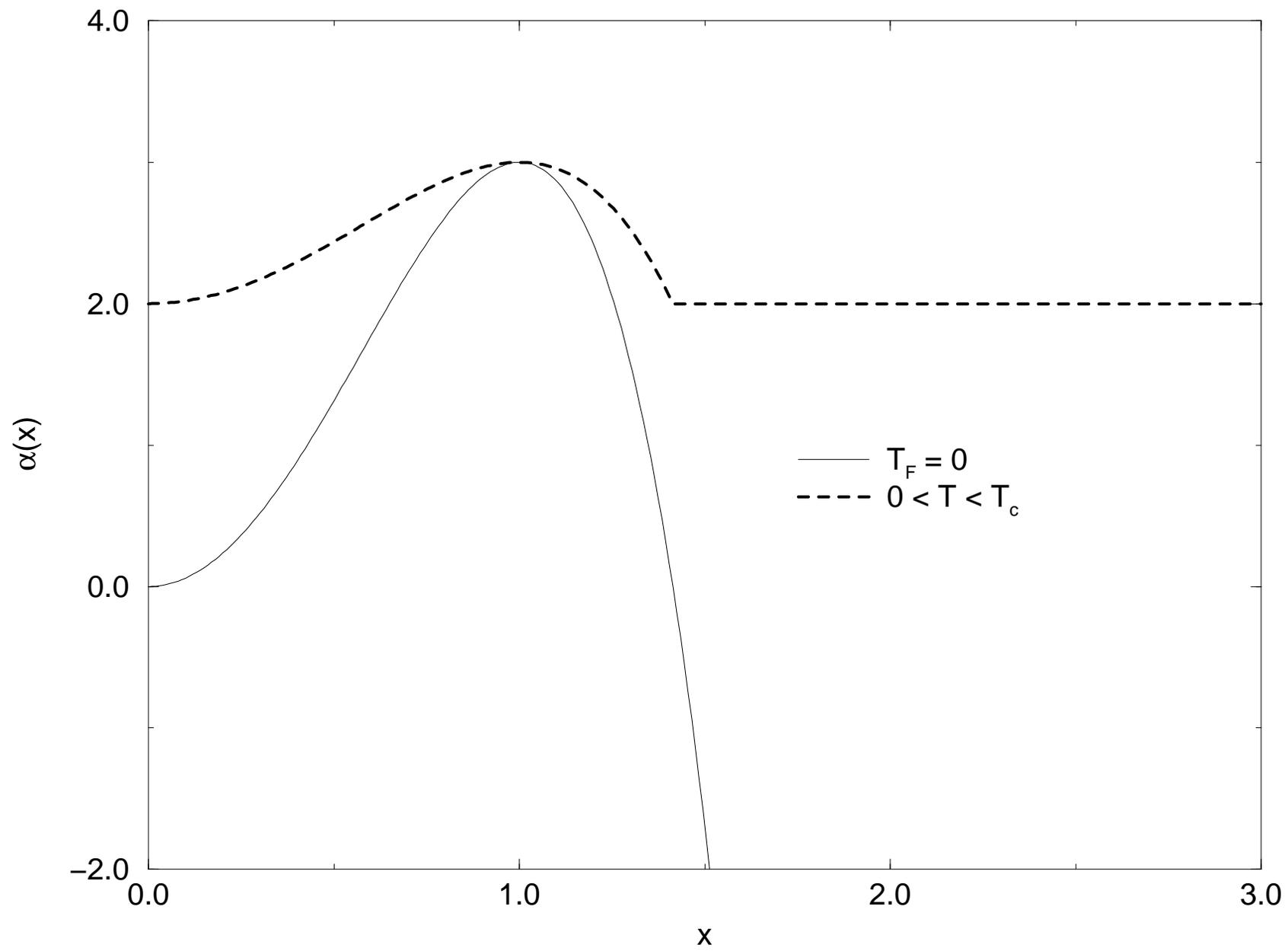
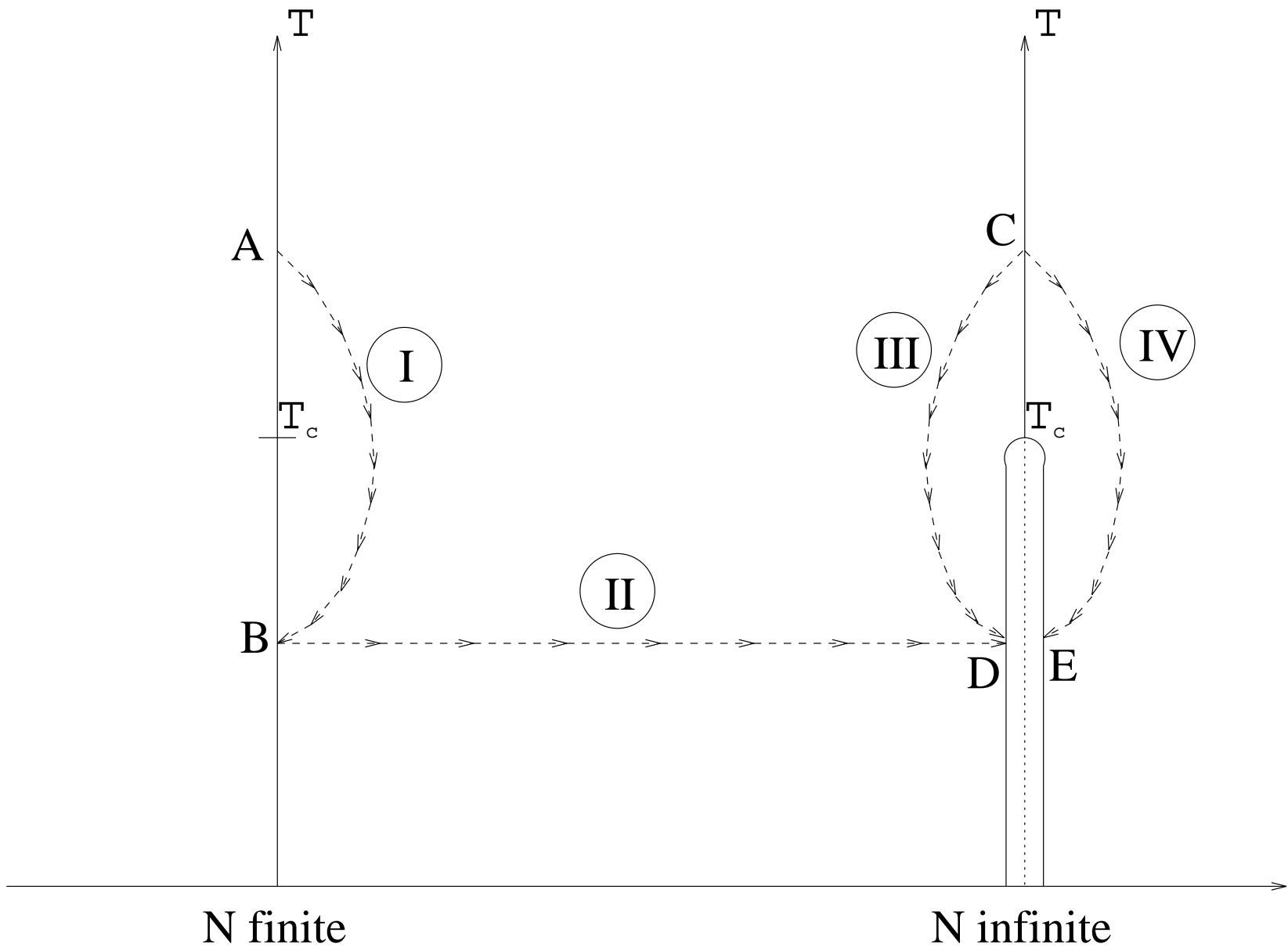
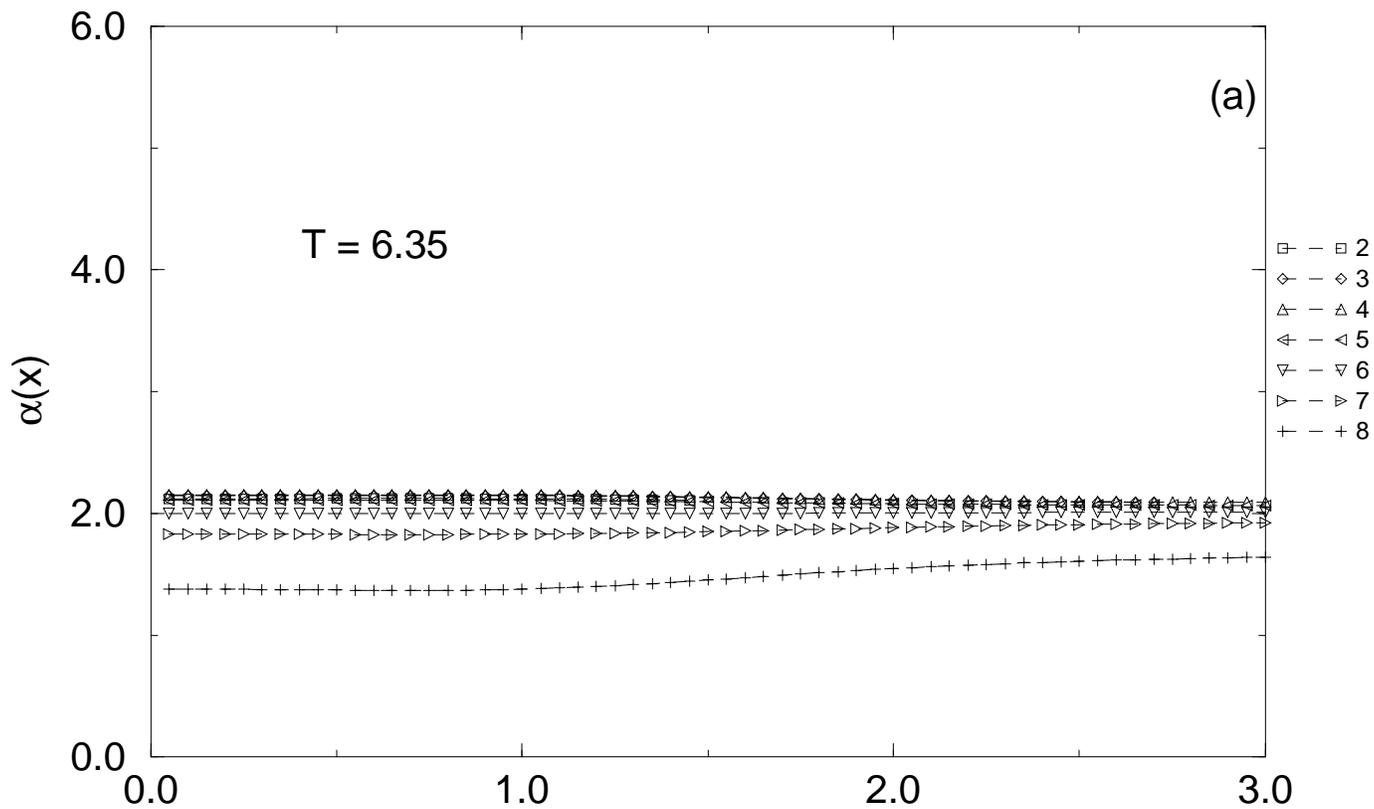
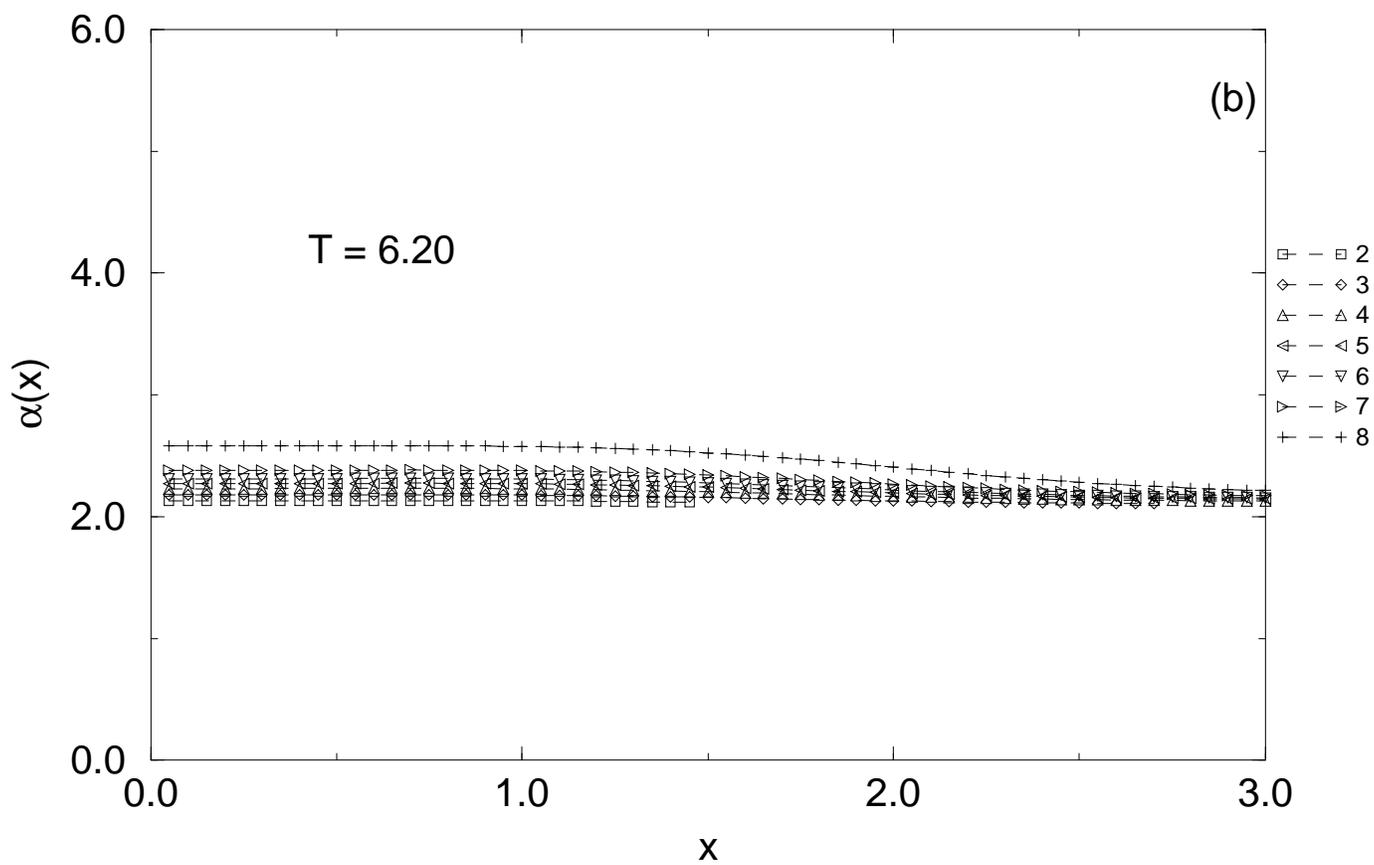
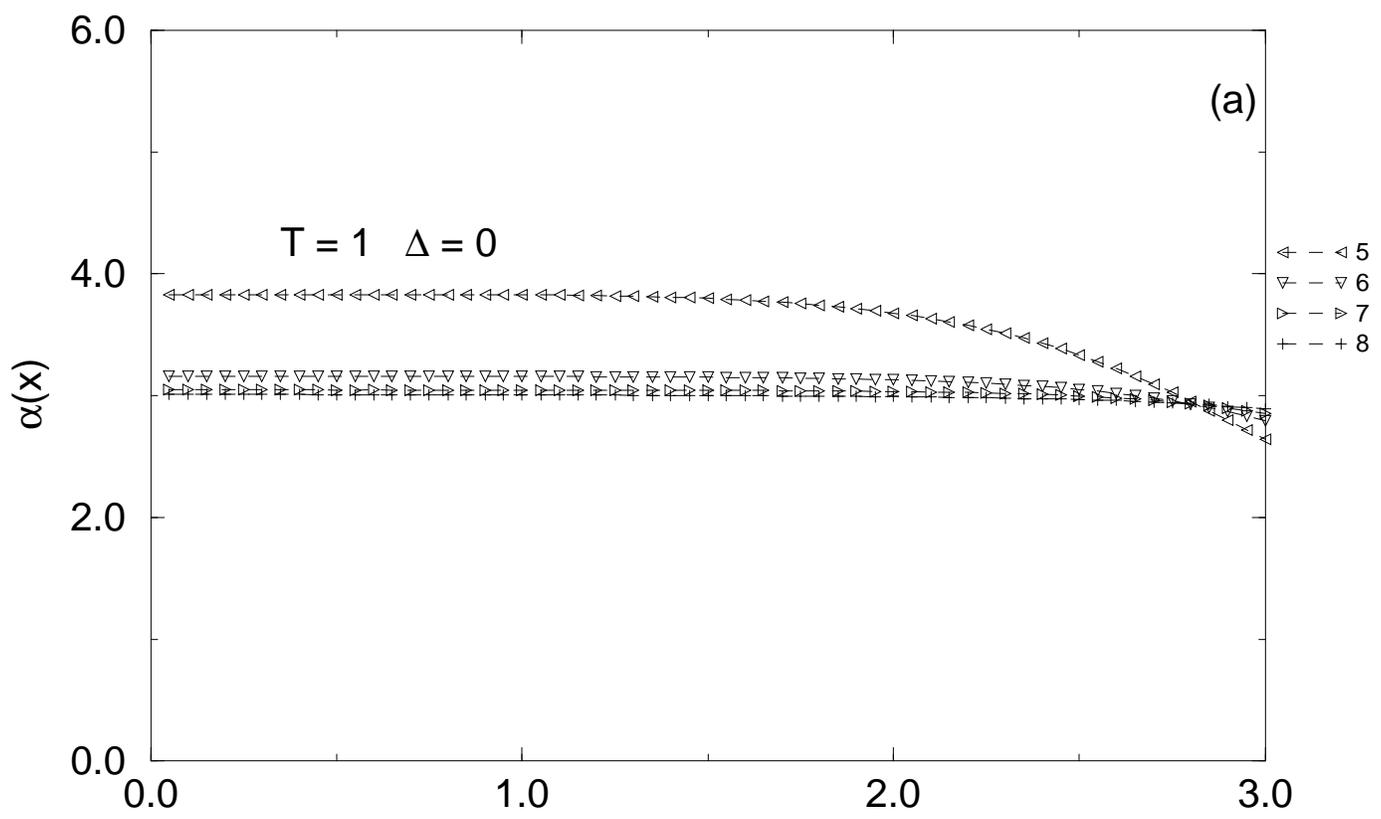
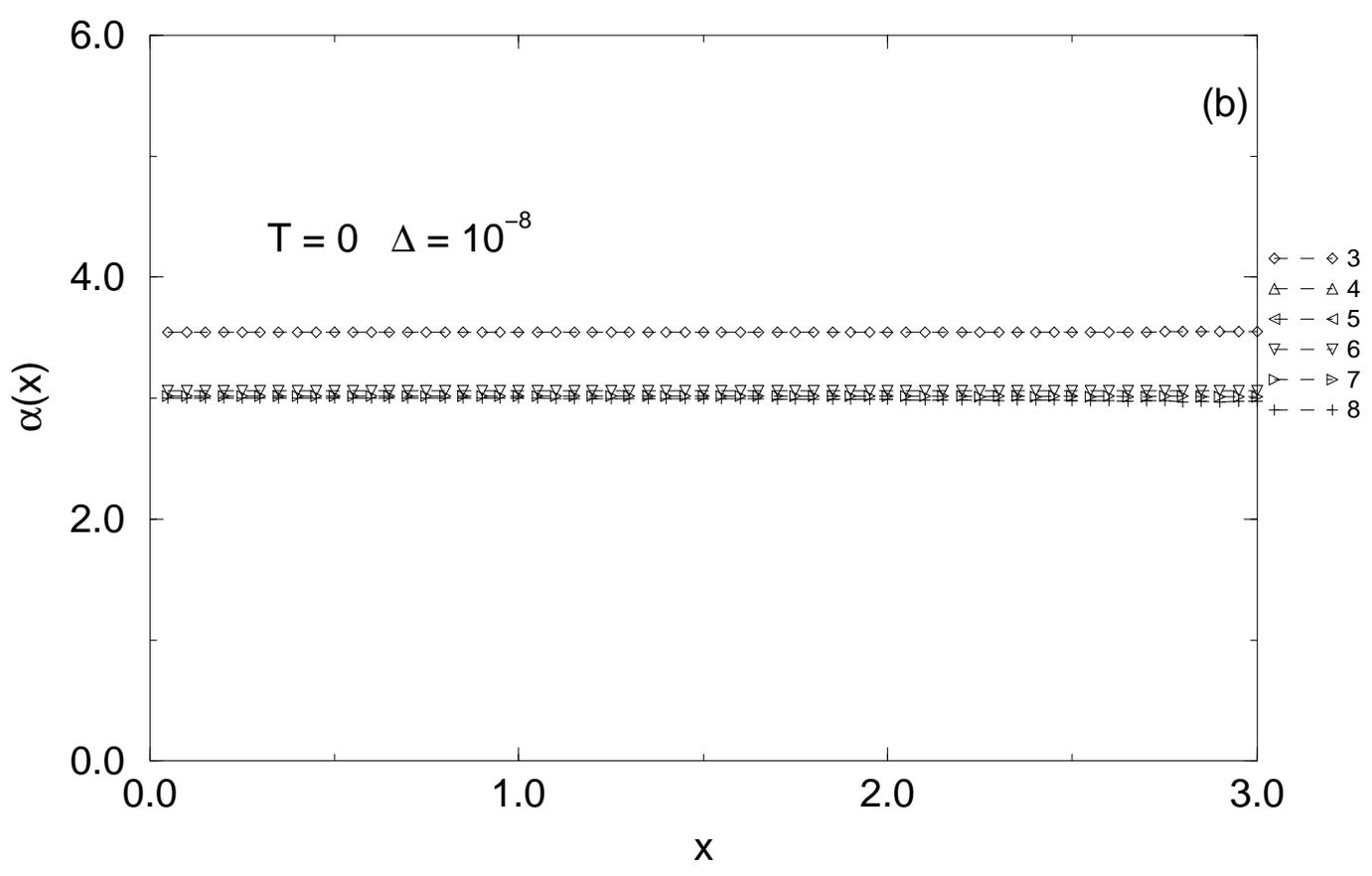
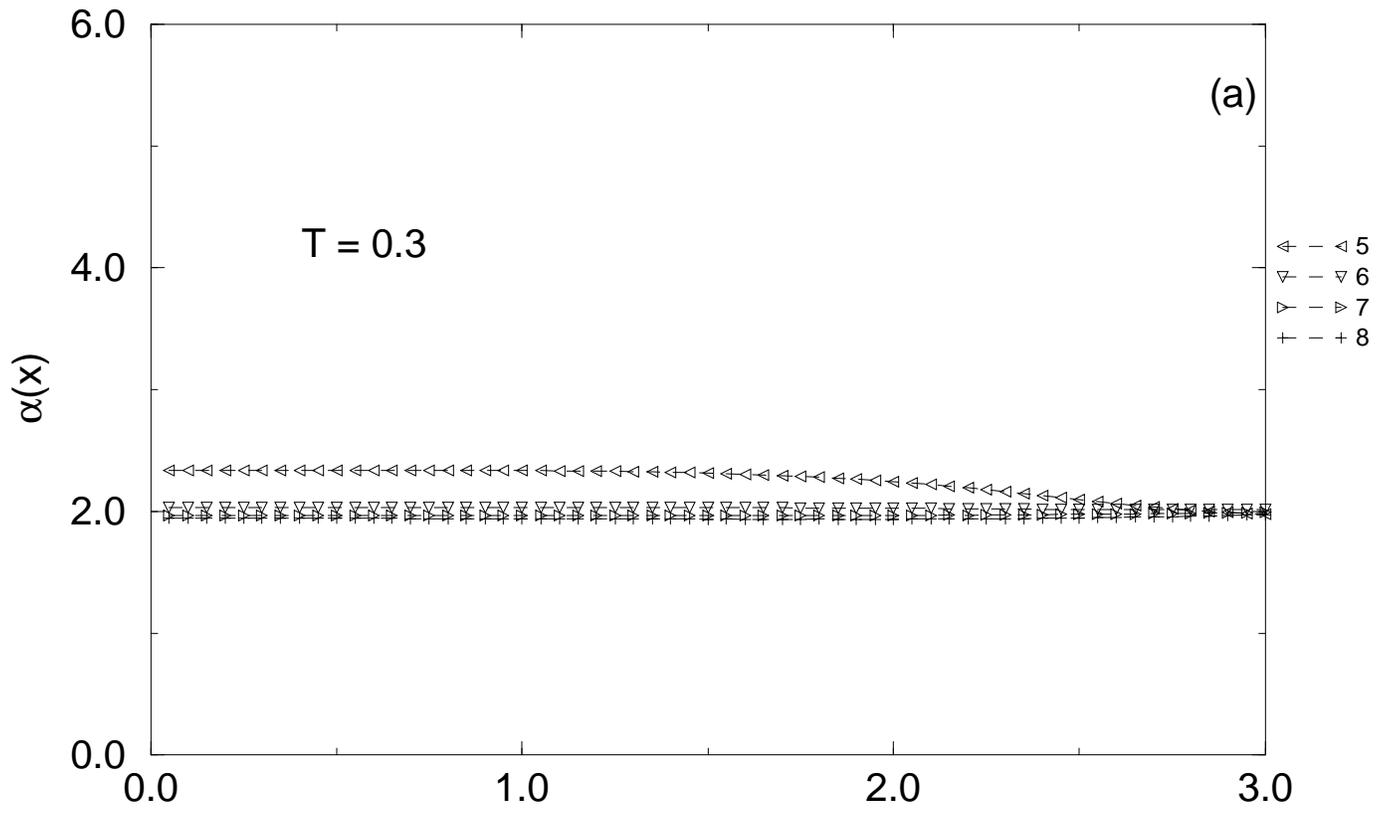
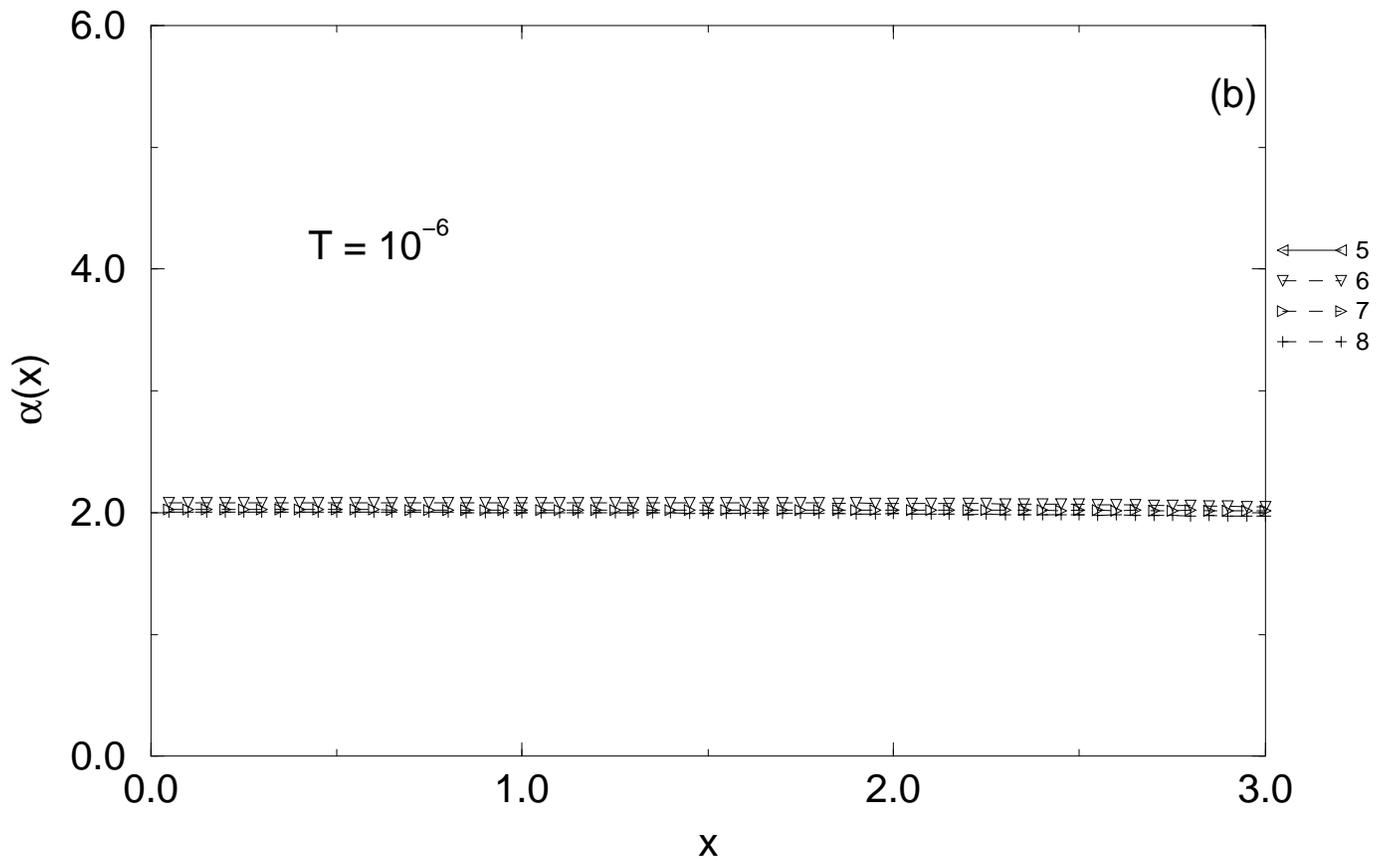
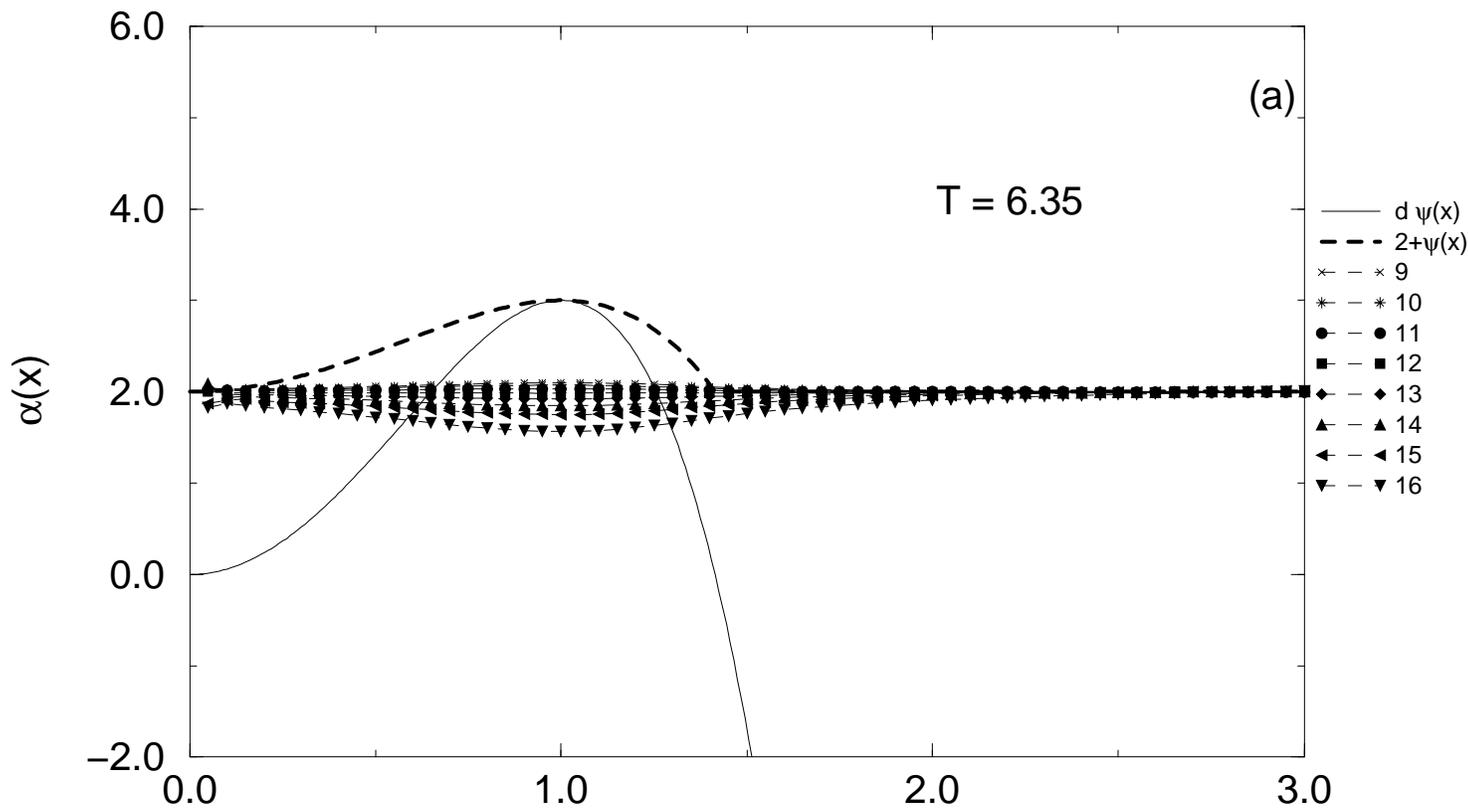
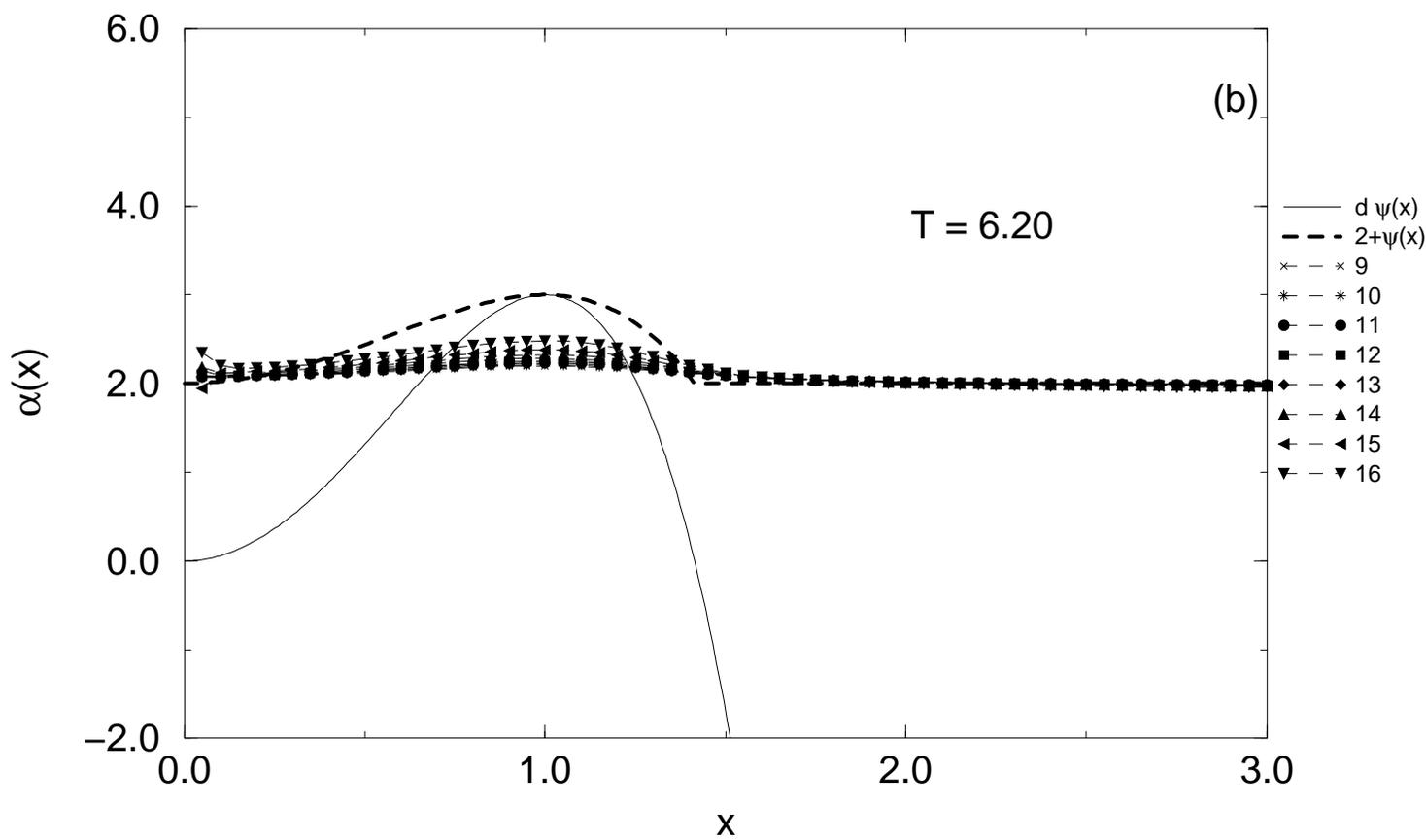
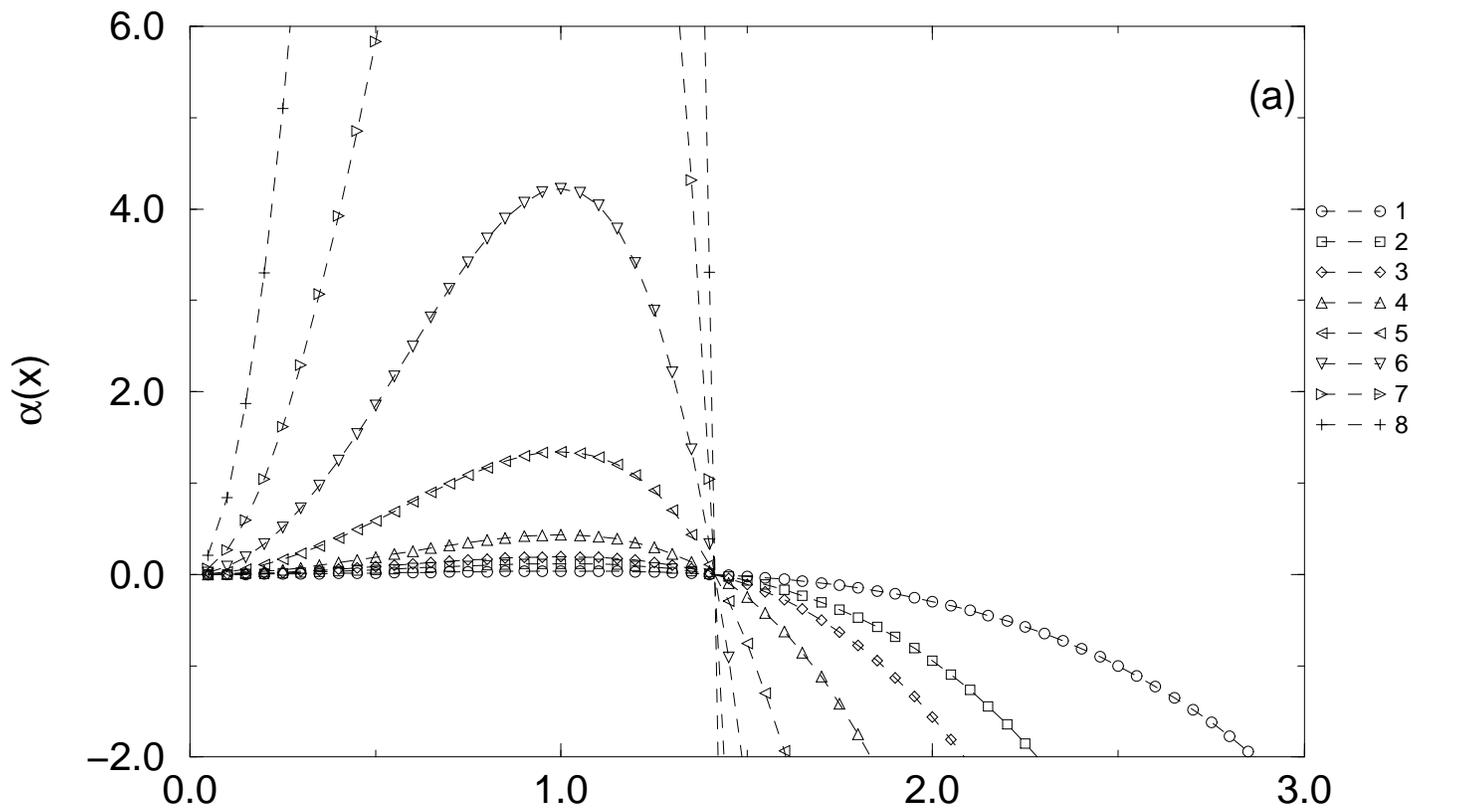
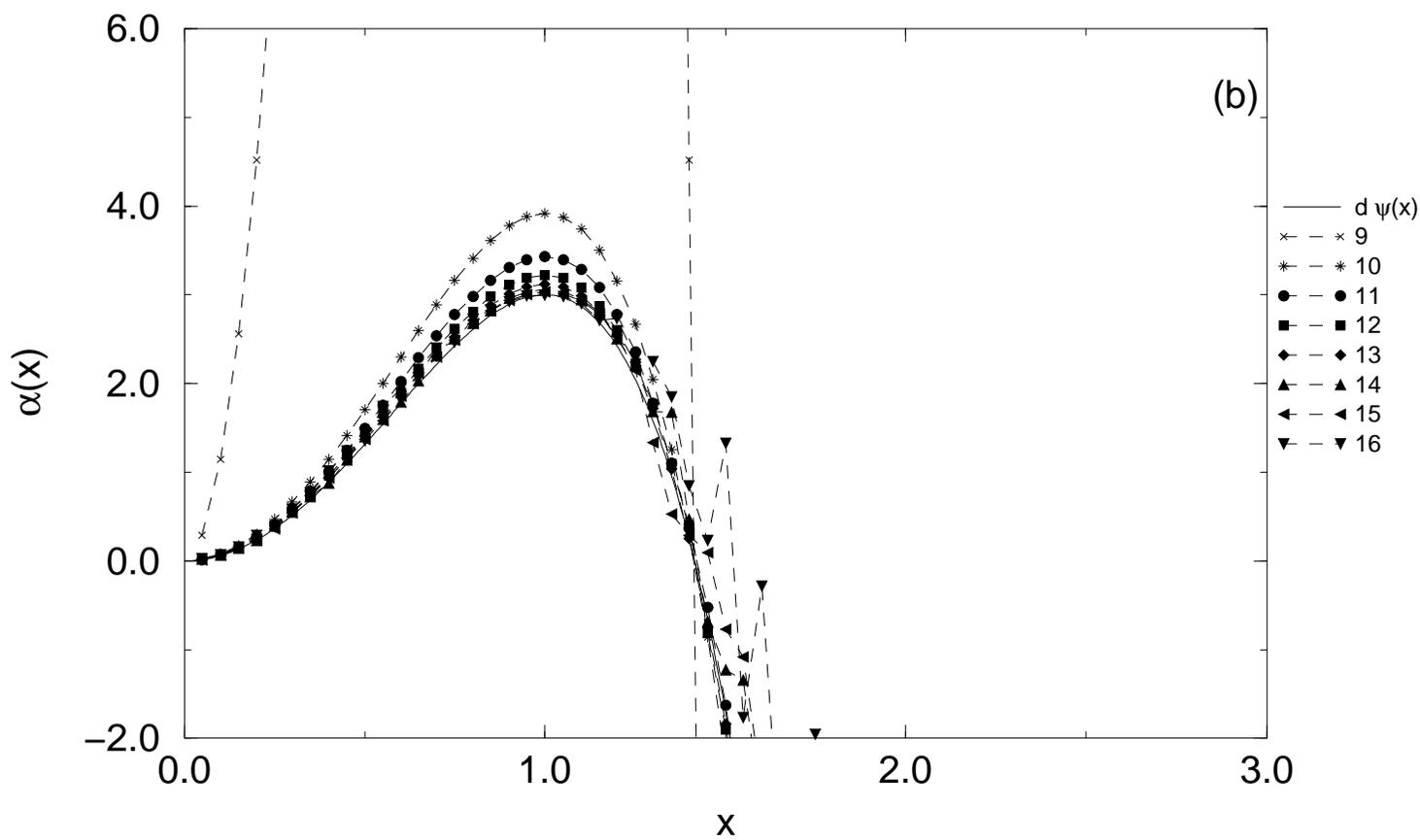
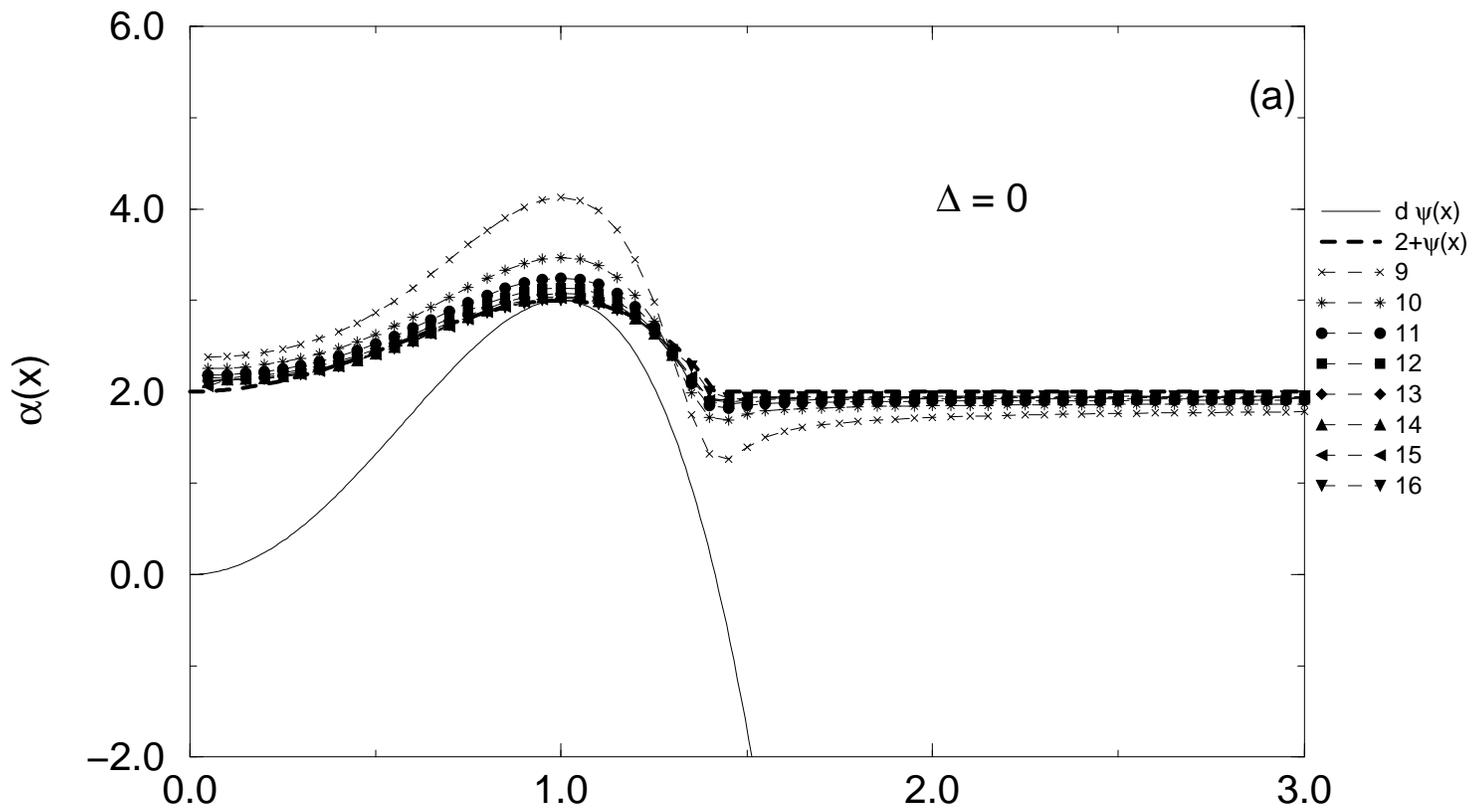
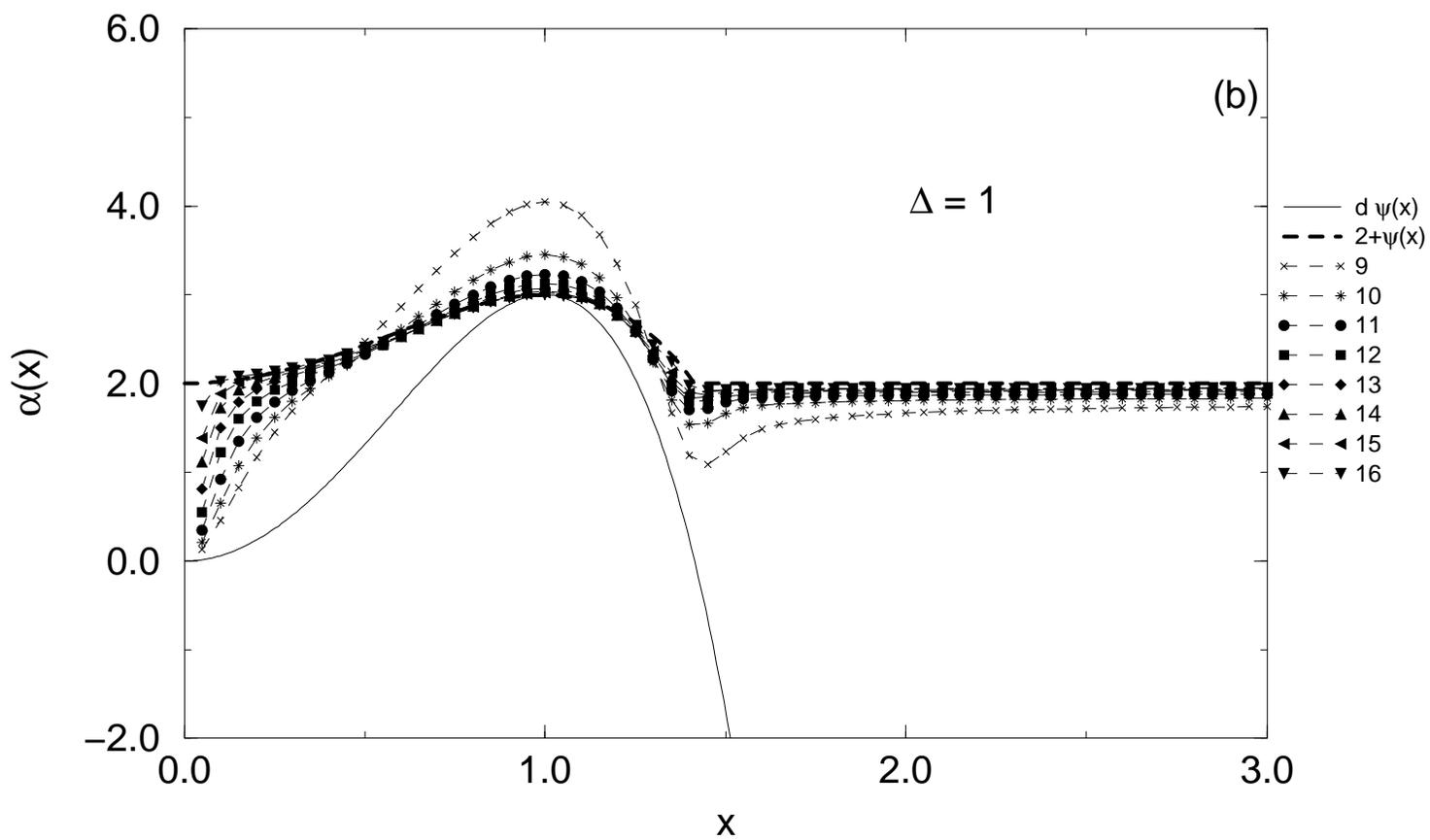
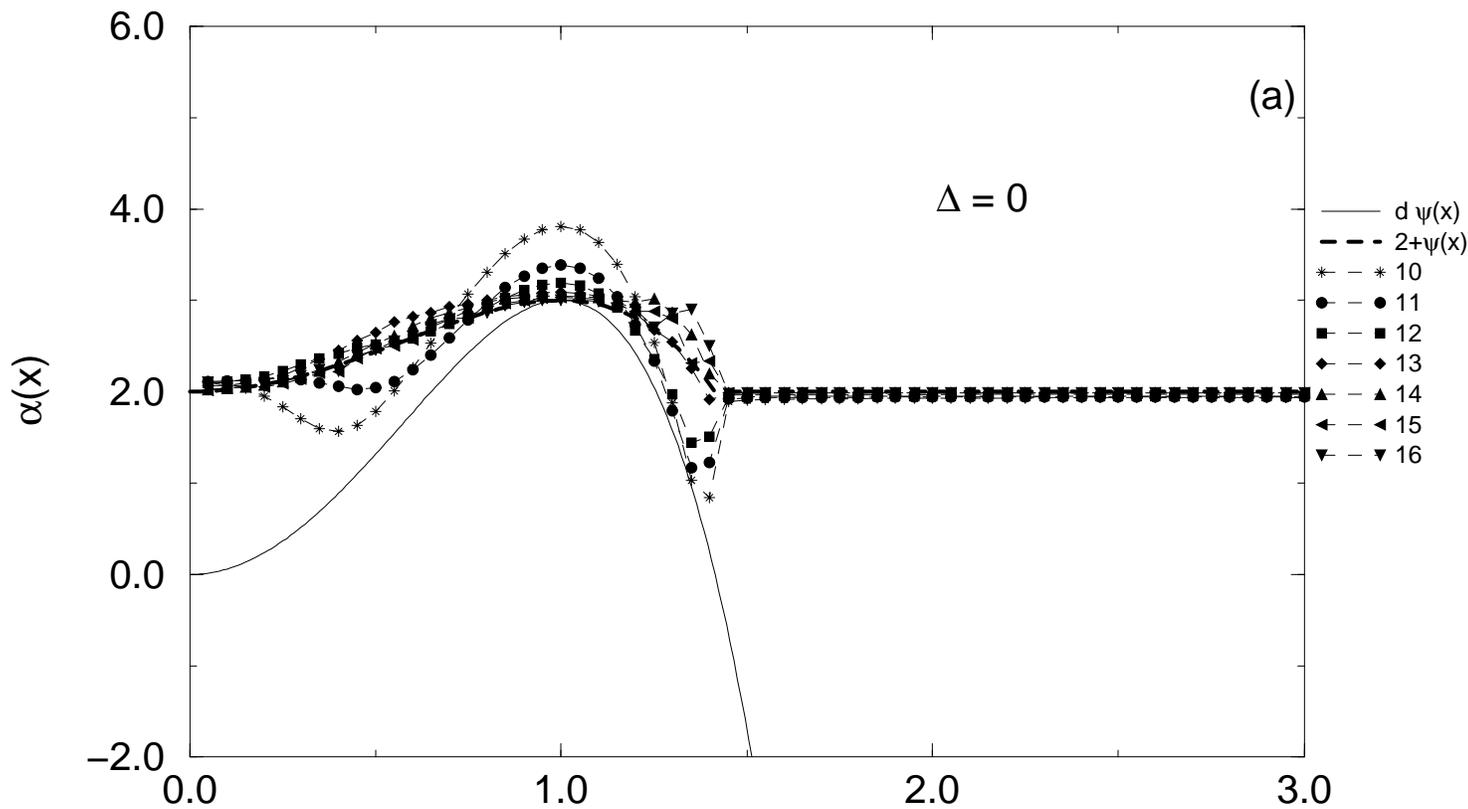
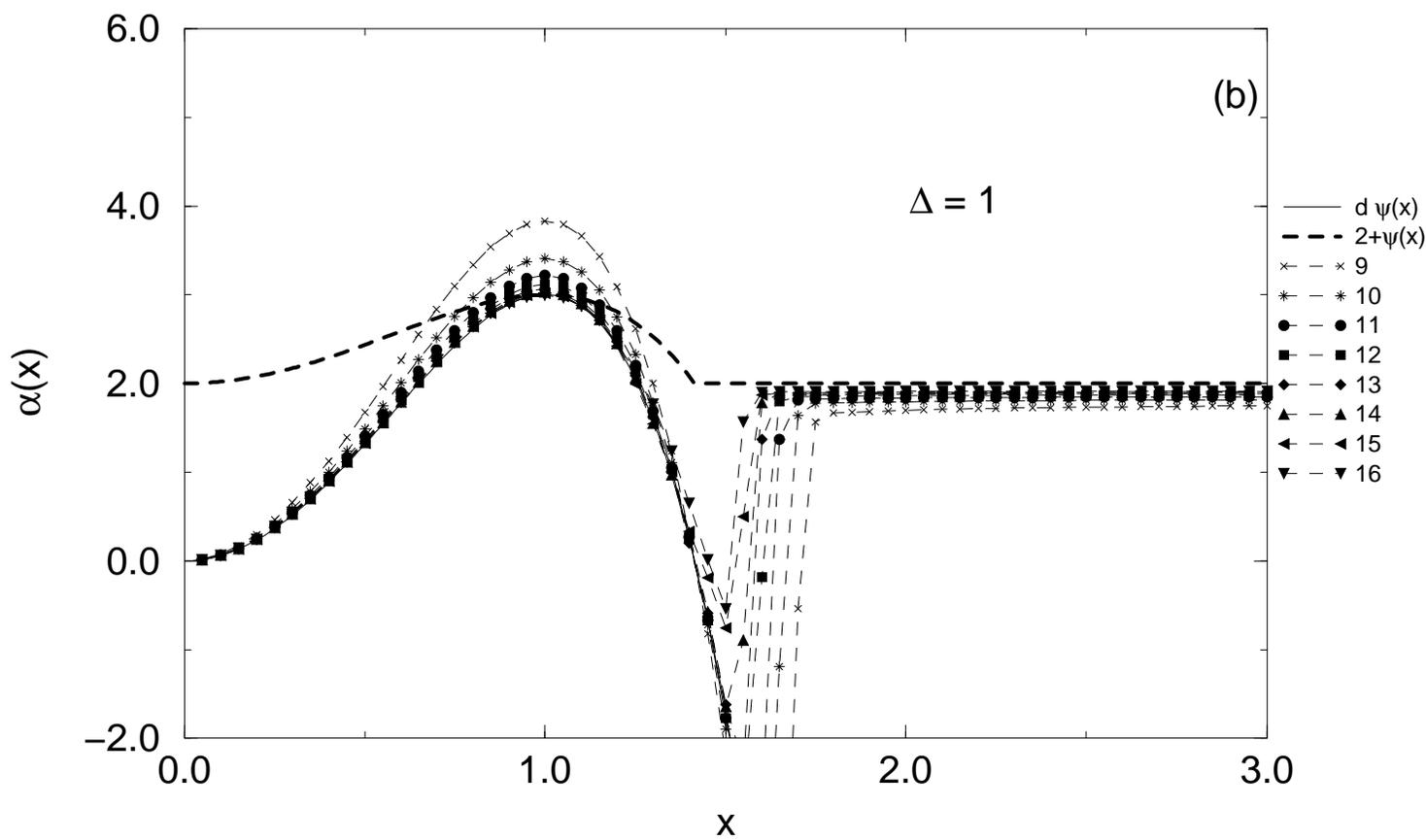
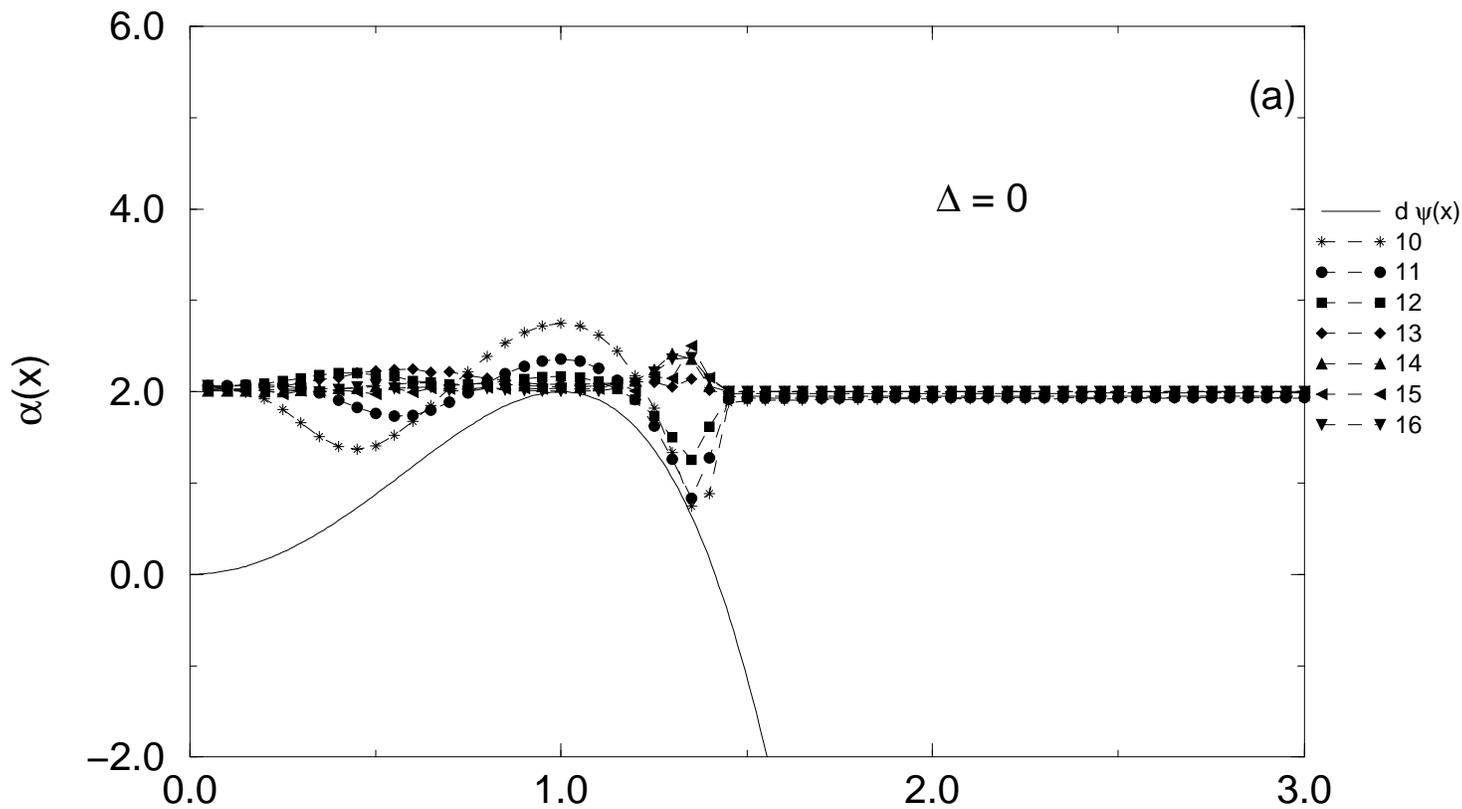
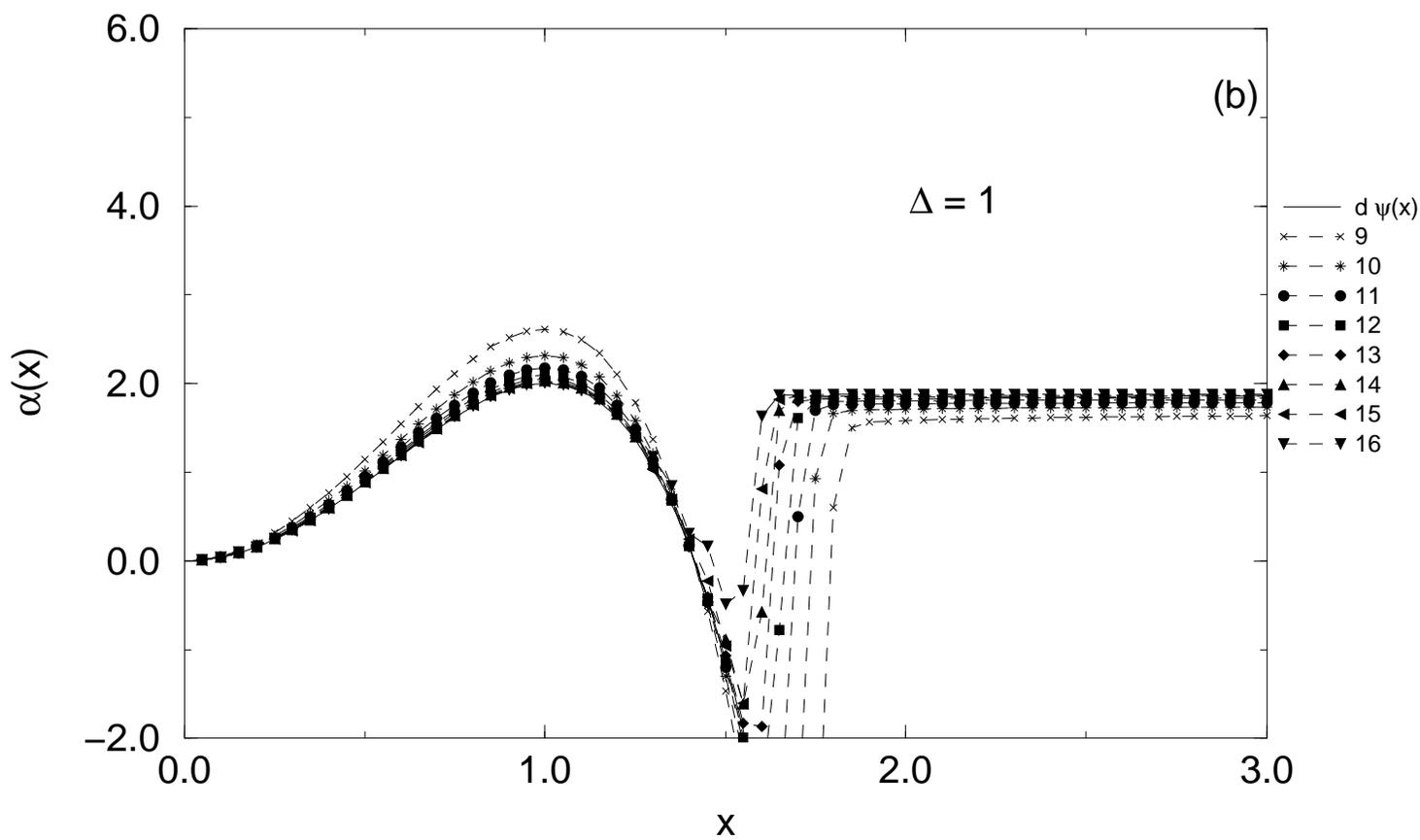
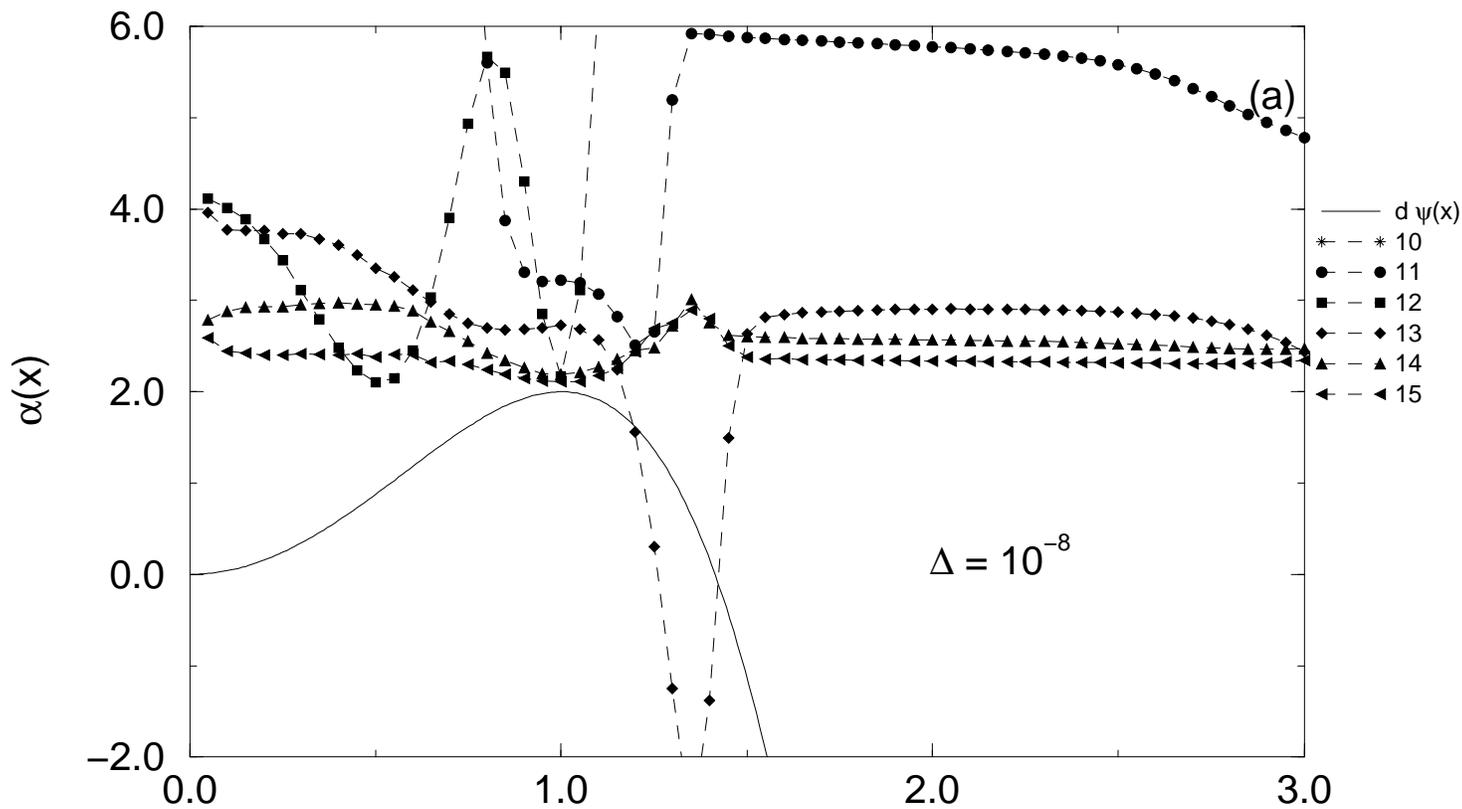
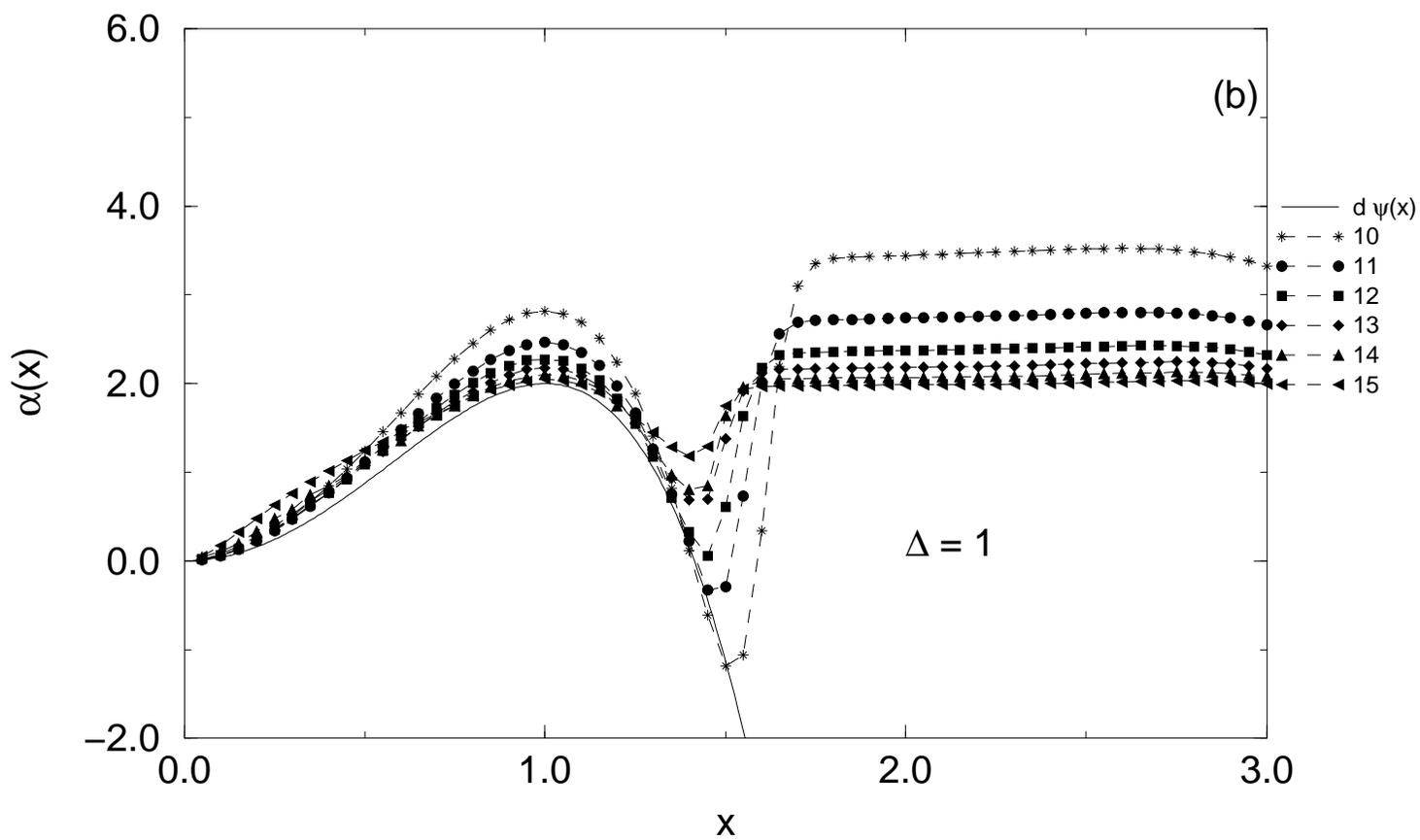
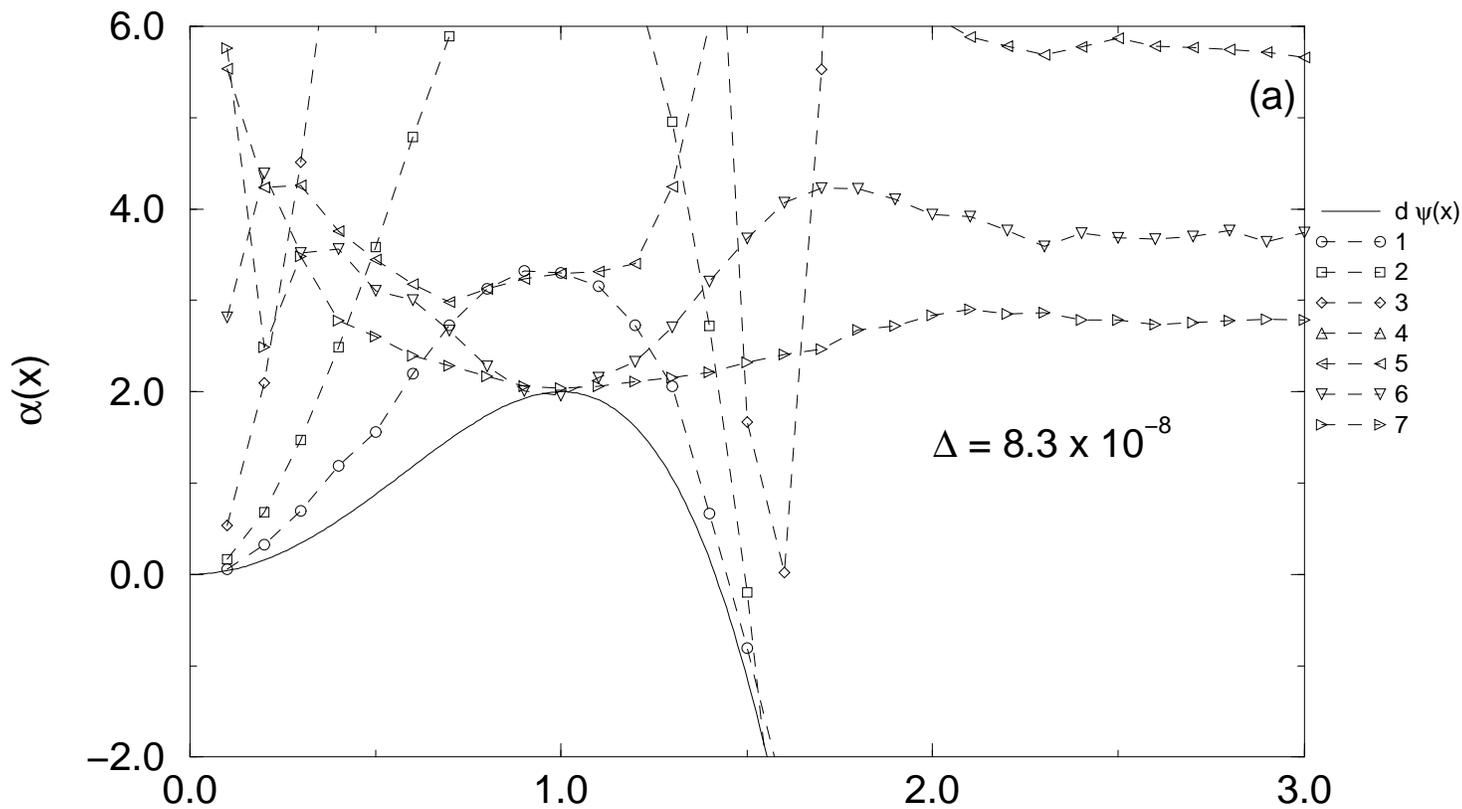
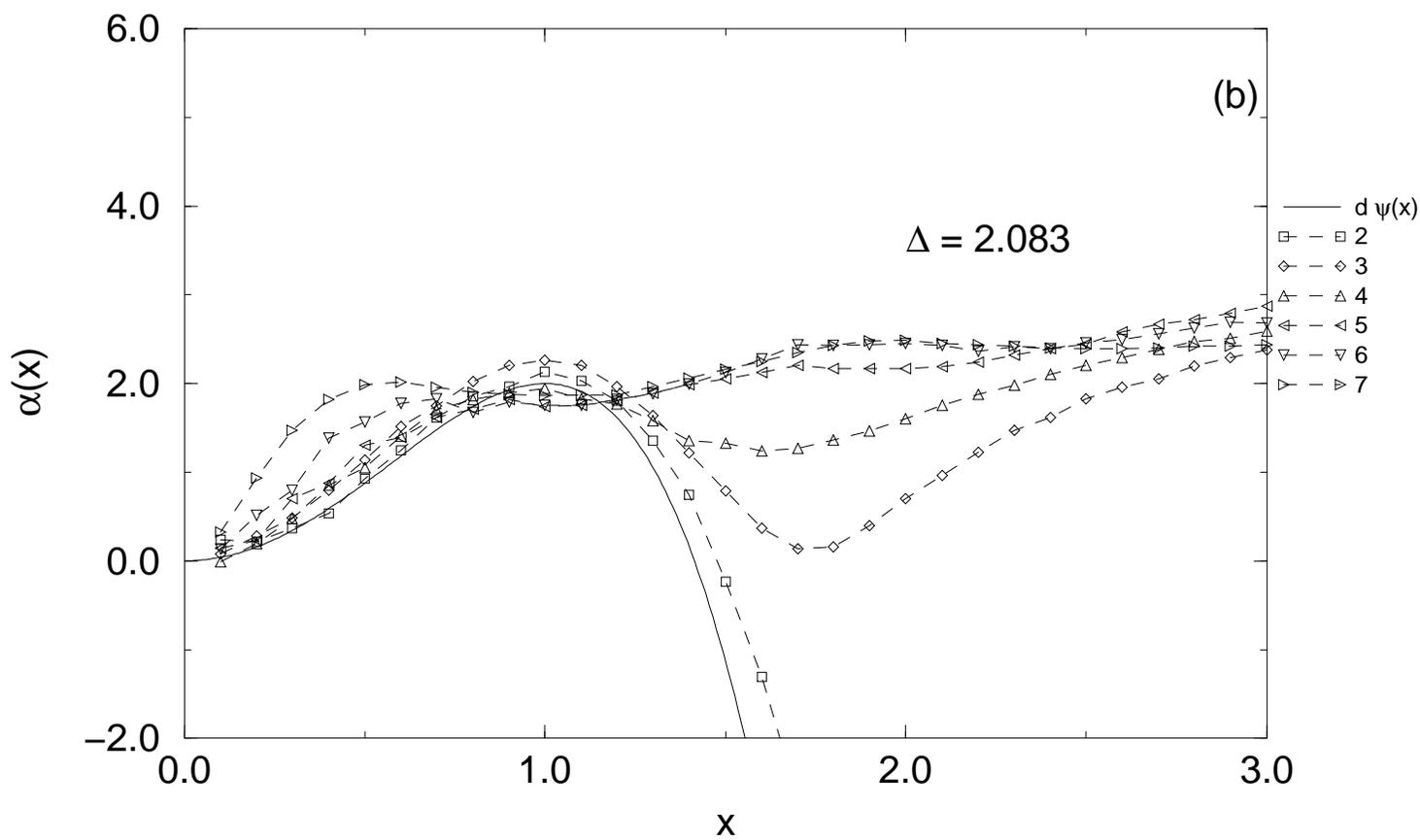
\end{document}